\documentclass{article}
\usepackage{arxiv}

\usepackage[utf8]{inputenc} 
\usepackage[T1]{fontenc}    
\usepackage{hyperref}       
\usepackage{url}            
\usepackage{booktabs}       
\usepackage{amsfonts}       
\usepackage{nicefrac}       
\usepackage{microtype}      
\usepackage{lipsum}
\usepackage{amsmath,mathtools,bbm,multirow,url,hyperref,enumerate}
\usepackage{color}
\usepackage{float}
\usepackage{hyperref}
\usepackage{makecell}
\usepackage{multicol}
\usepackage{natbib}
\hypersetup{colorlinks,citecolor=blue}

\usepackage{algorithm}
\usepackage{algpseudocode}

\usepackage{setspace}

\title{A Bayesian adaptive design for dual-agent phase I-II oncology trials integrating efficacy data across stages}

\author{
  Jos\'e L. Jim\'enez \\
  Novartis Pharma A.G.\\
  Basel, Switzerland\\
  \texttt{jose\_luis.jimenez@novartis.com}
    \And
  Haiyan Zheng \\
  MRC Biostatistics Unit, University of Cambridge\\
  Cambridge, UK\\
  \texttt{haiyan.zheng@mrc-bsu.cam.ac.uk}
}

\begin{document}
\maketitle

\begin{abstract}
Combination of several anti-cancer treatments has typically been presumed to have enhanced drug activity. Motivated by a real clinical trial, this paper considers phase I-II dose finding designs for dual-agent combinations, where one main objective is to characterize both the toxicity and efficacy profiles. We propose a two-stage Bayesian adaptive design that accommodates a change of patient population in-between. In stage I, we estimate a maximum tolerated dose combination using the escalation with overdose control (EWOC) principle. This is followed by a stage II, conducted in a new yet relevant patient population, to find the most efficacious dose combination. We implement a robust Bayesian hierarchical random-effects model to allow sharing of information on the efficacy across stages, assuming that the related parameters are either exchangeable or nonexchangeable. Under the assumption of exchangeability, a random-effects distribution is specified for the main effects parameters to capture uncertainty about the between-stage differences. The inclusion of nonexchangeability assumption further enables that the stage-specific efficacy parameters have their own priors. The proposed methodology is assessed with an extensive simulation study. Our results suggest a general improvement of the operating characteristics for the efficacy assessment, under a conservative assumption about the exchangeability of the parameters \textit{a priori}.
\end{abstract}

\keywords{Drug combination, information borrowing, meta-analytic-combined, phase I-II, seamless designs}

\maketitle

\section{Introduction}
\label{sc_introduction}

The primary objective of early phase clinical trials is to identify a dose that is safe and efficacious. Seamless phase I-II clinical trial designs are efficient approaches to study these two aspects in a single protocol. In the literature, we find two types of seamless phase I-II designs: one-stage and two-stage. The former usually estimates the joint probability of toxicity and efficacy using the accumulating data to recommend a best-suited dose to patients in the next cohort  \cite{ivanova2003new,thall2004dose,yuan2009bayesian,liu2018bayesian}. This setting is favoured when the efficacy outcome can be observed relatively soon after administration of the dose, e.g., after one or two cycles of therapy. By contrast, two-stage phase I-II designs come into play when efficacy cannot be ascertained in a short period of time. Specifically, stage I would commonly focus on toxicity considerations alone for dose (de-)escalation, despite that efficacy data are collected. This is then followed by stage II, where the evaluation of efficacy is the priority. A considerable amount of statistical literature has been written for the two-stage type \cite{rogatko2008patient,le2009dose,tighiouart2019two,jimenez2020bayesian,jimenez2021combining}.

Our work is motivated by the cisplatin-cabazitaxel clinical trial design proposed by Tighiouart (2019)\cite{tighiouart2019two}. In their proposed phase I-II study design, stage I was inspired by a conducted phase I trial \cite{lockhart2014phase} in patients with advanced solid tumors, where a single maximum tolerated dose (MTD) of cisplatin/cabazitaxel 15/75 $mg/m^2$ was recommended. Based on the available results and other preliminary efficacy data, it was hypothesized by the clinical team that there could be a series of tolerable and efficacious dose combinations for prostate cancer. Over the last years, clinical trials with drug combinations have received a fair amount of attention. This interest is motivated by the fact that drug combinations are able to induce synergistic treatment effects by simultaneously inhibiting resistance mechanisms and targeting multiple pathways. In stage II of the cisplatin-cabazitaxel trial \cite{tighiouart2019two}, 30 additional patients were enrolled to identify the dose combinations with high probability of efficacy, along the MTD curve estimated from stage I data. Another characteristic of the cisplatin-cabazitaxel trial \cite{tighiouart2019two} was that each stage has a different patient population, being the population in stage II the one of interest for the clinical team. Then, since the patient populations were not exactly the same in stages I and II, it was hypothesised that the dose-efficacy profiles could differ. Consequently, the dose-efficacy relationship was estimated using stage II data alone. The cisplatin-cabazitaxel trial \cite{tighiouart2019two} can further be improved in two ways: i) uncertainty about the estimated MTD curve should better be taken into account in stage II (i.e., the MTD curve may further be updated during stage II), and ii) efficacy data from stage I could be used for the final analysis by the end of stage II. The first limitation was recently addressed \cite{jimenez2021combining} allowing for a continuous update of the MTD curve throughout the entire phase I-II design. The novelty of the present article lies in addressing the second aspect; that is, we aim to integrate the efficacy information from both stages without neglecting the potential heterogeneity caused by the change of patient population across stages.

A robust Bayesian hierarchical model is fitted to allow combining the efficacy data across stages. The associated parameters are assumed to be either exchangeable or non-exchangeable. In our specific application, the benefit of borrowing is expected to lead to a precision improvement of the model parameter estimates and a reduction number of patients treated at sub-therapeutic dose combinations when there is a consistency between the efficacy profiles across stages. For cases of data inconsistency, the stage I efficacy data needs to be discounted effectively.

The manuscript is organized as follows. In Section \ref{sc_motivating_example}, we review the cisplatin-cabazitaxel trial, which serves as motivating example for the present work, as well as the proposed marginal dose-toxicity and dose-efficacy models to fit the trial. In Section \ref{sc_trial_design}, we introduce the proposed dose-finding algorithm for stages I and II, whereas in Section \ref{sc_simulation_study} we present a simulation study to evaluate the operating characteristics of the design, with focus on stage II. We provide concluding remarks in Section \ref{sc_discussion}.

\section{Motivating example and statistical models}
\label{sc_motivating_example}

\subsection{The cisplatin-cabazitaxel trial and data collection}

The original cisplatin-cabazitaxel trial enrolled patients with metastatic, castration resistant prostate cancer. A combination of continuous doses ranging from 10 to 25 $mg/m^2$ for cisplatin and from 50 to 100 $mg/m^2$ for cabazitaxel were administered intravenously every three weeks. As informed by a precedent study \cite{lockhart2014phase}, three specific combinations of cisplatin/cabazitaxel, 15/75, 20/75 and 25/75 $mg/m^2$, were evaluated. In stage I, the study enrolls 30 patients using conditional escalation with overdose control (EWOC) algorithm \cite{tighiouart2017bayesian} to estimate the MTD curve. In stage II, the study enrolls another 30 patients from the same population of patients but with visceral metastasis to identify dose combinations with high probability of efficacy along the MTD curve estimated at the end of stage I. These patients are allocated to dose combinations along the MTD curve using a Bayesian adaptive design after modeling the dose-efficacy curve with cubic splines.

The recommended MTD was 15/75 $mg/m^2$ on the basis of data from 24 patients (i.e., 9 evaluable patients in phase I and 15 patients in the expansion cohort) where only 2 out of 18 patients treated at the recommended MTD had dose limiting toxicity (DLT). Considering the low toxicity rate at the MTD reported \cite{lockhart2014phase}, as well as other (unpublished) preliminary efficacy data, the clinicians that contributed to the design of the Cisplatin-Cabazitaxel trial hypothesized that a series of tolerable dose combinations which could be efficacious in prostate cancer, could exist.

In this article, we regard the potential differences between stage I and stage II efficacy profiles from a different perspective. More specifically, we are motivated to establish a robust model formally accounting for such uncertainty, so that we can enhance the conduct and analysis of the stage II when there is a certain level of similarity between the efficacy profiles across the patient populations (the stages), as well as to discount the stage I efficacy data in case of dissimilarity.

\subsection{Problem formulation}
\label{sc_problem_formulation}

Let $x$ and $y$ be the respective dose levels, on their original continuous scales, of two compounds (labelled $X$ and $Y$) of interest, and further, $\{X_{\min}, Y_{\min}, X_{\max}, Y_{\max}\}$ be the lower and upper bounds. The measurement scales of $x$ and $y$ might differ from each other substantially. To avoid one variable being overly influential in the risk of toxicity, we standardise the doses using the transformations $h_1(x) = (x - X_{\min}) / (X_{\max} - X_{\min})$ and $h_2(y) = (y - Y_{\min}) / (Y_{\max} - Y_{\min})$, so that the standardised doses fall within the interval of [0, 1]. Thus, the dose combination $(0,0)$ corresponds to the lowest dose combination available in the trial and not to a lack of dose combination administration. For ease of notation, we retain the notation of $x$ and $y$ to denote the standardised dose levels.

Let $Z \in \{0,1\}$ be the binary indicator of DLT where $Z = 1$ represents the presence of a DLT and $Z = 0$ otherwise. Likewise, let $E \in \{0,1\}$ be the binary indicator of treatment response where $E=1$ represents a positive response, and $E=0$ otherwise. In this article, following the motivating trial, we assume that only the DLT can be observed rapidly after drug administration (e.g., after one cycle of therapy), whereas it takes three cycles or more for the efficacy outcome to be observable. Following \cite{tighiouart2019two}, let $\theta_T = 0.33$ be the target probability of DLT and $p_0 = 0.15$ be the probability of efficacy of the standard of care treatment. When employing synergistic cytotoxic agents, it is common to assume that both the dose-toxicity and dose-efficacy relationship are monotonically increasing functions. This implies that the optimal dose combination (i.e., the dose combination with most desirable benefit-risk trade-off) will lie in the MTD set, defined as $\mathcal{M} = \left \{ (x,y): P(Z=1|x,y) = \theta_T \right \}$, i.e., any dose combination $(x,y)$ with probability of DLT equal to $\theta_T$. A formal definition of the optimal dose combination is given in section \ref{sc_trial_design}. Given the two-stage formulation of this design, let $S \in \{1,2\}$ be the stage enrollment indicator to stage I and stage II, respectively, and let $\boldsymbol{D}_{S,i} = \{(Z_i, E_i, x_i, y_i)\}$ be the data collected in stage $S$ for the $i$-th patient.

\subsection{A marginal dose-toxicity model}

We assume that the binary outcomes of toxicity and efficacy are independent \cite{ivanova2009adaptive,cai2014bayesian,lyu2019aaa}. Alternatively, one could also account for the relationship between toxicity and efficacy either with the use of a Copula \cite{thall2004dose} or with a latent variable approach \cite{liu2018bayesian,lin2020adaptive}. However, this would add an additional layer of complexity to the design that is not in the scope of the article. Let the model for the marginal probability of DLT be
\begin{equation}
\label{modelprobdlt}
    \pi_T(x,y) = P(Z = 1|x,y) = F(\alpha_{0} + \alpha_{1}x + \alpha_{2}y + \alpha_{3}xy),
    \quad 
    \alpha_1,\alpha_2,\alpha_3 > 0,
\end{equation}where $F(.)$ is the cumulative distribution function of the logistic distribution, i.e., $F(u) = 1 / (1 + e^{-u})$. The parameters in this model can be interpreted as follows: i) $\alpha_0$ determines the probability of DLT at the lowest dose combination available in the trial, i.e., $(x=0, y =0)$, ii) $\alpha_1$ and $\alpha_2$ determine the contribution of compounds $X$ and $Y$ to the overall probability of DLT, and iii) $\alpha_3$ captures the potential increase in the probability of DLT due to drug-drug interaction.

Note that in Model (\ref{modelprobdlt}), because the number of attributable DLTs is expected to be very low given the cytotoxic nature of cisplatin and cabazitaxel, we do not take into account toxicity attributions \cite{jimenez2019cancer}.

We reparameterize the marginal probability of DLT defined in Model \eqref{modelprobdlt} in terms of parameters that clinicians can easily interpret \cite{tighiouart2017bayesian}. Let $\rho_{uv}$ denote the joint probability of DLT when the levels of agents $X = u$ and $Y = v$, with $u \in \{0, 1\}$, and $v \in \{0, 1\}$, so that $\alpha_0 = F^{-1}(\rho_{00})$, $\alpha_1 = (F^{-1}(\rho_{10}) - F^{-1}(\rho_{00}))$, and $\alpha_2 = (F^{-1}(\rho_{01}) - F^{-1}(\rho_{00}))$. The MTD thus has the form of

\begin{equation}
\label{mtdcurve}
\begin{split}
& \mathcal{M} = \left \{ (x,y): y = \frac{(F^{-1}(\theta_T) - F^{-1}(\rho_{00})) - (F^{-1}(\rho_{10}) - F^{-1}(\rho_{00})) x}{(F^{-1}(\rho_{01}) - F^{-1}(\rho_{00})) + \alpha_3 x } \right \}. 
\end{split}
\end{equation}

Following Tighiouart (2019) \cite{tighiouart2019two}, we use informative prior distributions based on the results of Lockhart et al. (2014) \cite{lockhart2014phase} so that $\rho_{10}, \rho_{01}$ and $\alpha_3$ are independent \emph{a priori} with $\rho_{01} \sim \mbox{Beta}(1.4,5.6)$, $\rho_{10} \sim \mbox{Beta}(1.4,5.6)$, and conditional on $(\rho_{01}, \rho_{10})$, $\rho_{00} / \min (\rho_{01}, \rho_{10}) \sim \mbox{Beta}(0.8, 7.2)$. Also, let the interaction parameter $\alpha_3 \sim \mbox{Gamma(0.8, 0.0384)}$, i.e., a shape of 0.8 and a rate of 0.0384. These prior distributions imply that the combination of cisplatin/cabazitaxel 15/75 $mg/m^2$ has a probability of DLT approximately equal to 0.33. The posterior distribution of the dose-toxicity model parameters is defined as $p({\boldsymbol \rho}, \alpha_3 | \boldsymbol{D}_{1}) \propto p({\boldsymbol \rho}, \alpha_3) \times \mathcal{L}(\boldsymbol{D}_{1} | {\boldsymbol \rho}, \alpha_3)$, where ${\boldsymbol \rho} = \{\rho_{00}, \rho_{01}, \rho_{10}\}$ and where $\boldsymbol{D}_1$ corresponds to the data from stage 1.

\subsection{A marginal dose-efficacy model}
\label{sc_marginal_efficacy}

We now shift our focus to estimate the dose-efficacy relationship in a dual-dimensional plane; for stage $S = \{1, 2\}$, we stipulate the stagewise dose-efficacy data model as
\begin{equation}
\label{eq_dose_efficacy_model}
    \pi_E^{S}(x,y) = P(E = 1|x,y,S) = F(\beta_{0S} + \exp(\beta_{1S})x + \exp(\beta_{2S})y + \beta_{3S}xy),
\end{equation}where $F(.)$ remains to be the cumulative distribution function of the logistic distribution. Because the motivating trial employs cytotoxic agents, we assume that the probability of efficacy does not decrease with the dose of any agent when the other agent is held constant. To ensure this property, we apply the exponential function to $\beta_{1S}$ and $\beta_{2S}$ since $\exp(u) > 0$. We also assume that $\beta_{3S} > 0$, which means that there is a synergistic effect due to the interaction of the two compounds \cite{gasparini2013general}. The parameters in this model can be interpreted as follows: i) $\beta_0$ determines the probability of efficacy at the lowest dose combination available in the trial, i.e., $(x=0, y =0)$, ii) $\beta_1$ and $\beta_2$ determine the contribution of compounds $X$ and $Y$ to the overall probability of efficacy, and iii) $\beta_3$ captures the potential increase in the probability of efficacy due to drug-drug interaction.

Let $\boldsymbol{\Psi}_S = (\beta_{1S}, \beta_{2S})$ denote the main effects of the treatment specific to stage. We consider a \emph{meta-analytic-combined} (MAC) approach \cite{neuenschwander2016use} to establish a Bayesian predictive distribution for $\boldsymbol{\Psi}_2 \mid \boldsymbol{D}_1, \boldsymbol{D}_2$. This would allow the investigator to estimate main effects of drugs X and Y, using the efficacy data from both stages.

We assume a normal-normal hierarchical model to relate the stagewise main effects of efficacy for the dual-agent. Specifically, at stage $S = 1$
$$
\boldsymbol{\Psi}_1 \mid \boldsymbol{\mu}, \boldsymbol{\Phi} \sim \mbox{BVN}(\boldsymbol{\mu}, \boldsymbol{\Phi}).
$$
Continuing the phase I-II trial to stage $S = 2$ in a new population, we introduce a non-exchangeability distribution and stipulate that 
\begin{equation}
\label{eq_bvn}
\begin{split}
\boldsymbol{\Psi}_2 \mid \boldsymbol{\mu}, \boldsymbol{\Phi} & \sim \mbox{BVN}(\boldsymbol{\mu}, \boldsymbol{\Phi})   \qquad \text{ with probability } \omega, \\
\boldsymbol{\Psi}_2 & \sim \mbox{BVN}(\boldsymbol{m}_0, R_0) \qquad \text{ with probability } 1 - \omega,
\end{split}
\end{equation}with 
\begin{equation}
\label{eq_bvn_distribution_ex}
{\boldsymbol \mu} = \begin{pmatrix}
\mu_{1}\\ 
\mu_{2}
\end{pmatrix}, \qquad {\boldsymbol \Phi} = \begin{pmatrix}
\tau_1^2 & \xi \tau_1 \tau_2\\ 
\xi \tau_1 \tau_2 & \tau_2^2
\end{pmatrix},
\end{equation}and
\begin{equation}
 \textbf{m}_0 = \begin{pmatrix}
0\\ 
0
\end{pmatrix}, \qquad \textbf{R}_0 = \begin{pmatrix}
10^2 & \zeta  100\\ 
\zeta 100 & 10^2
\end{pmatrix}.
\end{equation}

The variance terms in ${\boldsymbol \Phi}$ represent between-stage heterogeneity. Mind that in each stage of the cisplatin-cabazitaxel trial we have two distinct and well designated populations. Consequently, between-stage heterogeneity and between-population heterogeneity cannot be disentangled. If our trial would involve multiple populations in each stage, we would need additional random-effects distributions to account for the between-population differences\cite{zheng2020robust,zheng2021bridging}. The values of \textbf{m}$_0$ and \textbf{R}$_0$ are selected so that they induce weakly informative prior distributions over the parameters in $\boldsymbol{\Psi}_2$. This Bayesian hierarchical random-effects model is completed by the following hyperpriors:
\begin{equation*}
\mu_1 \sim N(\eta_1,s_1^2), \quad \mu_2 \sim N(\eta_2,s_2^2), \quad \tau_1 \sim \text{HN}(z_1), \quad \tau_2 \sim \text{HN}(z_2), \quad \xi \sim U(0,0.5), \quad \zeta \sim U(0,0.5),
\end{equation*}where HN($z$) denotes a half-normal distribution formed by truncating a $N(0,z^2)$ so it covers the interval $(0,\infty)$. We select HN(0.5) anticipating for substantial between-stage heterogeneity in the main effects model parameters. Other viable choices of HN($z$) and the indication have been noted\cite{zheng2020robust, roychoudhury2020bayesian}. The value $z = 0.5$ serves as a weakly informative prior distribution, although the values of $z_1$ and $z_2$ can be justified appropriately for the user's own case, with evidence suggesting the similarity or dissimilarity of efficacy in such two patient populations.

The specification of $\omega$ requires, in practice, the input of subject-matter experts and needs to be fixed \emph{a priori}. We place weakly informative prior distributions over the dose-efficacy model parameters:
\begin{equation*}
\begin{split}
& \beta_{01} \sim N(-1.8, 3.16^2), \quad \beta_{02} \sim N(-1.8, 3.16^2), \\
&\beta_{31} \sim \text{Gamma}(0.1, 0.1), \quad \beta_{32} \sim \text{Gamma}(0.1, 0.1).
\end{split}
\end{equation*}For illustration purposes, we set
\begin{equation*}
\begin{split}
& \eta_1 = 0, \quad s_1 = 3.16, \quad \eta_2 = 0, \quad s_2 = 3.16, \quad z_1 = 0.5, \quad z_2 = 0.5,
\end{split}
\end{equation*}to implement the model. Overall, the weakly informative prior distributions we select in this article translate into the median probability of efficacy estimates with 95\% credible intervals displayed in Table S1 of the supplementary material.


\section{An integrated phase I-II design for dose finding}
\label{sc_trial_design}

Stage I will enroll a total of $N_1 = C_1 \times m_1$ patients, where $C_1$ denotes the total number of cohorts in phase I with each of the size $m_1$. Stage II will enroll a total of $N_2 = n_2 + C_2 \times m_2$ patients, where $n_2$ is the number of patients in the first cohort of stage II, $C_2$ the additional number of cohorts and $m_2$ its size. 

In the original cisplatin-cabazitaxel trial, stage I efficacy data was entirely discarded and therefore, at the beginning of stage II, an initial cohort $n_2$ was used to collect efficacy data homogeneously across the entire MTD curve. In this article, we choose to keep $n_2$ as a short run-in period that can inform the data (in)consistency and thus determine the degree of information sharing.

Let $N = N_1 + N_2$ be the total number of patients that the entire study will enroll and $\widehat{\mathcal{M}}_{\boldsymbol{D}_1}$ be the estimated MTD set based on data from stage I. We select $m_1 = 2$, $m_2 = 5$ and $n_2 = 10$. At the end of stage II, we test the following null and alternative hypotheses
\begin{equation}
\label{eq_hypotheses}
\begin{split}
& H_0: \pi_E^{S=2}(x,y) \leq p_0 \text{ for all  } (x,y) \in \widehat{\mathcal{M}}_{\boldsymbol{D}_1} \quad \mbox{vs.} \\
&H_1: \pi_E^{S=2}(x,y) > p_0 \text{ for some  } (x,y) \in \widehat{\mathcal{M}}_{\boldsymbol{D}_1}. \\
\end{split}
\end{equation}and we reject the null hypothesis if
\begin{equation}
    \underset{(x,y) \in \widehat{\mathcal{M}}_{\boldsymbol{D}_1}}{\mbox{arg max}} \left ( P(\pi_E^{S=2}(x,y) > p_0 | \boldsymbol{D}_1, \boldsymbol{D}_2) \right ) > \delta_u,
    \label{eq_optimal_dose_comb}
\end{equation}where $\delta_u = 0.4$ is a pre-specified design parameter. Moreover, the dose combination
\begin{equation}
\label{opt_dose}
(x,y)_{opt}=\underset{(x,y) \in \widehat{\mathcal{M}}_{\boldsymbol{D}_1}}{\mbox{argmax}} P(\pi_E^{S=2}(x,y) > p_0 | \boldsymbol{D}_1, \boldsymbol{D}_2), 
\end{equation}is recommended as the optimal dose combination and is selected for further phase IIb or III studies.

As previously mentioned, stage I is based on the escalation with overdose control (EWOC) principle \cite{babb1998cancer, tighiouart2005flexible, tighiouart2010dose, tighiouart2017bayesian, tighiouart2012number, shi2013escalation} where the posterior probability of overdosing the next cohort of patients is bounded by a feasibility bound $\alpha$. For the definition of the algorithm, let $\lambda(\Gamma_{X|Y=y} | \boldsymbol{D}_1)$ represent the posterior distribution of the MTD of drug $X$ given that the level of drug $Y$ is equal to $y$ (i.e., given that $Y$ is fixed) based on stage I data $\boldsymbol{D}_1$ (see equation \eqref{mtdcurve} for the definition of the MTD). Also, let $\Lambda_{\Gamma_{X | Y = y}}^{-1}(\alpha | \boldsymbol{D}_1)$ denote the $\alpha$-th percentile of $\lambda(\Gamma_{X|Y=y} | \boldsymbol{D}_1)$. In a cohort with two patients, the first one would receive a new dose of compound $X$ given that the dose $y$ of compound $Y$ that was previously assigned. The other patient would receive a new dose of compound $Y$ given that dose $x$ of compound $X$ was previously assigned. These steps are described in Stage I of Algorithm \ref{algorithm1}. Using EWOC, these new doses are at the $\alpha$-th percentile of the conditional posterior distribution of the maximum tolerated dose combinations. The feasibility bound $\alpha$ increases from 0.25 up to 0.5 in increments of 0.05 (see Wheeler et al. (2017)\cite{wheeler2017toxicity}). Accrual continues until the maximum sample size in stage I is reached or the trial is stopped early for safety.

Stage II follows the response-adaptive randomization principle. This type of Monte Carlo algorithm uses the current parameter estimates to sample a cohort of $m_2$ dose combinations from the estimated dose-efficacy standardized density of $\widehat{\pi}_E^{S=2} (x,y)$ along the estimated MTD curve. Note that $\widehat{\pi}_E^{S=2} (x,y)$ uses the Bayes estimates of the dose-efficacy model parameters. As explained in Jim\'enez and Tighiouart (2022) \cite{jimenez2021combining}, because stage II selects doses on the estimated MTD curve $\widehat{\mathcal{M}}_{\boldsymbol{D}_1}$, and there is a one-to-one correspondence between $(x,y) \in \widehat{\mathcal{M}}_{\boldsymbol{D}_1}$, we may write $\widehat{\pi}_E^{S=2} (x,y) = \widehat{\pi}_E^{S=2} (x)$ for $(x,y) \in \widehat{\mathcal{M}}_{\boldsymbol{D}_1}$. In other words, by having the value of $x$, using the definition of the MTD in equation \eqref{mtdcurve} we can easily obtain the corresponding value of $y$. Thus, to facilitate the definition of the standardize density function, instead of writing $\widehat{\pi}_E^{S=2} (x,y)$ we simply write $\widehat{\pi}_E^{S=2} (x)$. The standardized density of the estimated efficacy curve is $\Tilde{\pi}_E^{S=2} (x) = \frac{\widehat{\pi}_E^{S=2} (x)}{\int_{x \in X' \widehat{\pi}_E^{S=2} (x) dx}}$. A rejection sampling algorithm is then used to sample $m_{2}$ dose combinations from this density. These steps are are described in Stage II of Algorithm \ref{algorithm1}.

\begin{algorithm}[H]
\caption{Stage I and stage II algorithms}
\begin{algorithmic}
\State \textbf{STAGE I}
\State - In the first cohort ($c_1 = 1$) patients 1 and 2 receive the dose combination $(x_1, y_1) = (x_2, y_2) = (0.33, 0.5)$
\State - In the second cohort ($c_1 = 2$) patients 3 and 4 receive doses $(x_3, y_3)$ and $(x_4, y_4)$ respectively, where $y_3 = y_1$, $x_4 = x_2$, $x_3$ is the $\alpha$-th percentile of $\lambda(\Gamma_{X|Y = y_1} | \boldsymbol{D}_{1,2})$, $y_4$ is the $\alpha$-th percentile of $\lambda(\Gamma_{Y | X = x_2} | \boldsymbol{D}_{1,2})$.
\For{$c_1 = 3:C_1$}
  \If{$c_1$ is an even number}
   \State - Patient $2 c_{1} - 1$ receives the dose combination $(x_{2c_{1}-1}, y_{2c_{1}-3})$ where $x_{2c_{1}-1} = \Lambda_{\Gamma_{X | Y = y_{2c_{1}-3}}}^{-1}(\alpha | \boldsymbol{D}_{1,2c_{1}-2})$
   \State - Patient $2c_{1}$ the dose combination $(x_{2c_{1}-2}, y_{2c_{1}})$, where $y_{2c_{1}} = \Lambda_{\Gamma_{Y | X = x_{2c_{1}-2}}}^{-1}(\alpha | \boldsymbol{D}_{1,2c_{1}-2})$
   \Else
   \State - Patient $2c_{1} - 1$ receives the dose combination $(x_{2c_{1} - 3}, y_{2c_{1} - 1})$ where $y_{2c_{1} - 1} = \Lambda_{\Gamma_{Y | X = x_{2c_{1}-3}}}^{-1}(\alpha | \boldsymbol{D}_{1,2c_{1}-2})$
   \State - Patient $2c_{1}$ receives the dose combination $(x_{2c_{1}}, y_{2c_{1} - 2})$ where $x_{2c_{1}} = \Lambda_{\Gamma_{X | Y = y_{2c_{1}-2}}}^{-1}(\alpha | \boldsymbol{D}_{1,2c_{1}-2})$
  \EndIf
 \EndFor
 \State \textbf{STAGE II} 
 \State - Calculate the posterior median of the parameters $\rho_{00}, \rho_{10}, \rho_{01}$ and $\alpha_3$ given data ${\boldsymbol{D}}_1$, i.e., $(\widehat{\rho}_{00}, \widehat{\rho}_{10}, \widehat{\rho}_{01}, \widehat{\alpha}_3)$.
 \State - Calculate estimated MTD set $\widehat{\mathcal{M}}_{\boldsymbol{D}_1} = \left \{ (x,y): y = \left ( \frac{F^{-1}(\theta_Z) - F^{-1}(\widehat{\rho}_{00}) - (F^{-1}(\widehat{\rho}_{10}) - F^{-1}(\widehat{\rho}_{00})) x}{(F^{-1}(\widehat{\rho}_{01}) - F^{-1}(\widehat{\rho}_{00})) + \widehat{\alpha}_3 x} \right ) \right \}$
 \State - Allocate $n_2$ patients to dose combinations equally spaced along the estimated MTD curve $\widehat{\mathcal{M}}_{\boldsymbol{D}_1}$.
 \State - Calculate the posterior median of the parameters $(\widehat{\beta}_{02}, \widehat{\beta}_{12}, \widehat{\beta}_{22}, \widehat{\beta}_{32})$ using the MAC approach given data ${\boldsymbol D}_1,{\boldsymbol D}_2$.
 \For{$c_2 = 1:C_2$}
    \State - Generate a sample of dose combinations of size $m_2$ that belong to $\widehat{\mathcal{M}}_{\boldsymbol{D}_1}$ from the (estimated) standardized density $\widehat{\pi}_E^{S=2}(x,y)$, and assign it to the subsequent cohort of $m_2$ patients.
    \State - Calculate the posterior median of the parameters $(\widehat{\beta}_{02}, \widehat{\beta}_{12}, \widehat{\beta}_{22}, \widehat{\beta}_{32})$ using the MAC approach given data ${\boldsymbol D}_1,{\boldsymbol D}_2$.
\EndFor 
\end{algorithmic}
\label{algorithm1}
\end{algorithm}

The dose finding algorithm contains the following stopping rules for safety and futility:

\begin{itemize}
    \item \textbf{Futility stopping rule}:

        For ethical considerations and to avoid exposing patients to sub-therapeutic dose combinations, we would stop the trial for futility if 
        \begin{equation*}
            \underset{(x,y) \in \widehat{\mathcal{M}}_{\boldsymbol{D}_1}}{\mbox{arg max }}\left ( \pi_E^{S=2}(x,y)  > p_0 | \boldsymbol{D}_1, \boldsymbol{D}_2) \right ) < \delta_0,
        \end{equation*}where $\delta_0$ is a pre-specified threshold. For the purposes of illustration in this article, we choose $\delta_0 = 0.1$. Mind that this stopping rule applies only after the run-in cohort of $n_2$ patients in stage II.

\item \textbf{Safety stopping rule}

The design contains two stopping rules for safety, one for stage I and a less stringent one for stage II. During stage I, we would stop the trial if
\begin{equation}
\label{eq_tox_stop_st1}
   P(\pi_T(x = 0,y = 0) > (\theta_T + 0.1) ~|~ \boldsymbol{D}_1) > \delta_{\theta_1},
\end{equation}where $\delta_{\theta_1} = 0.5$. In contrast, during stage II we would stop the trial if
\begin{equation}
\label{eq_tox_stop_st2}
   P(\Theta > (\theta_T + 0.1) ~|~ \boldsymbol{D}_2) > \delta_{\theta_2},
\end{equation} where $\Theta$ represents the rate of DLTs for both stages of the design regardless of dose and $\delta_{\theta_2} = 0.9$ represents the confidence level (i.e., 90\%) that a prospective trial results in an excessive DLT rate. A non-informative Jeffrey's prior Beta$(0.5,0.5)$ is placed on the parameter $\Theta$.
\end{itemize}

\section{Simulation study}
\label{sc_simulation_study}

\subsection{Operating characteristics}
\label{sc_operating_characteristics}

In this section, we present a simulation study that will assess the operating characteristics of our design. Since we apply an already established dose-escalation procedure in stage I, we concentrate on evaluating the design's operating characteristics for the stage II, which leverages efficacy data from stage I. We report the simulation results according to the following metrics:

\begin{itemize}
    \item Distribution of the recommended optimal dose combinations.
    
    \item Proportion of recommended optimal dose combinations with true probability of efficacy above $p_0$. For simplicity, this metric is referred as the \textit{percentage of correct recommendation}.
    
    \item (Approximated) Bayesian power (or type-I error probability under $H_0$):
    \begin{equation}
    \label{eq_power_formula}
    \mbox{Power} \approx \frac{1}{J} \sum_{j = 1}^{J} \textbf{1} \left \{ \underset{(x,y) \in \widehat{\mathcal{M}}_{\boldsymbol{D}_1}}{\mbox{max }} \left ( P \left (\pi_{E,j}^{S=2}(x,y) > p_0 | \boldsymbol{D}_1, \boldsymbol{D}_2  \right ) \right )   > \delta_u \right \},
    \end{equation}where `$\textbf{1}(.)$' represents an indicator function and $J$ represents the total number of simulated trials, and $\pi_{E,j}^{S=2} = F \left (\beta_{02}^{(j)} + \exp \left (\beta_{12}^{(j)} \right ) x + \exp \left (\beta_{22}^{(j)} \right ) y + \beta_{32}^{(j)} xy \right )$. Under null scenarios as defined in \eqref{eq_hypotheses}, the above formula represents the (approximated) Bayesian type-I error probability.
    
    \item  Average posterior probability of early stopping for futility and safety.
    
    \item  Proportion of patients in stage II allocated to dose combinations with true probability of efficacy above $p_0$.
   
\end{itemize}

\subsection{Scenarios}
\label{sc_scenarios}

In stage I, we construct two dose-toxicity scenarios considered as highly plausible by the principal investigator of the motivating trial \cite{tighiouart2019two,jimenez2021combining}. The true dose-toxicity model parameters are presented in Table S2 in the supplementary material, and displayed in Figure \ref{Figure_1}. Furthermore, we assume that stage II has the same dose-toxicity profile as stage I (i.e., the dose-toxicity profiles does not vary across patient populations)\cite{tighiouart2019two,jimenez2020bayesian,jiang2021shotgun}. The target probability of DLT is $\theta_T = 0.33$.

\begin{figure}[H]
\caption{MTD curves obtained with the dose-toxicity model parameter values presented in Table S2 in the supplementary material. The point at the cisplatin/cabazitaxel 15/75 $mg/m^2$ combination represents the MTD found by Lockhart et al. (2014)\cite{lockhart2014phase}.}
\centering
\vspace{0.25cm}
\includegraphics[scale=0.6]{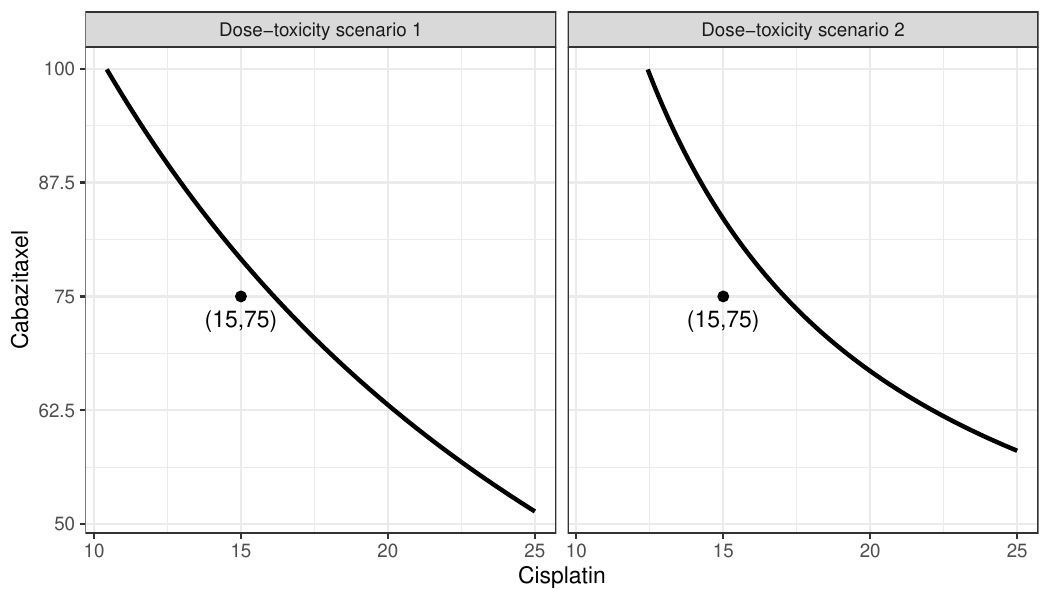}
\label{Figure_1}
\end{figure}

For each of the dose-toxicity scenarios, we consider two different stage II dose-efficacy profiles that place the dose combination with highest efficacy in opposite locations. In terms of the stage I dose-efficacy profiles, we consider the following three hypothetical situations:

\begin{enumerate}
    \item The stage I and stage II dose-efficacy profiles are perfectly consistent. For reading purposes, we refer to this profile as ``complete agreement between stage I and stage II dose-efficacy profiles'' or simply as ``CA'', which is short for ``Complete Agreement''.
    
    \item The stage I and stage II dose-efficacy profiles point to the same dose combination with highest efficacy, but the probabilities of efficacy are different across stages. For reading purposes, we refer to this profile as ``partial agreement between stage I and stage II dose-efficacy profiles'' or simply as ``PA'', which is short for ``Partial Agreement''.
    
    \item The stage I and stage II dose-efficacy profiles are completely different, and place the dose combination with highest efficacy in different locations. For reading purposes, we refer to this profile as  ``complete disagreement between stage I and stage II dose-efficacy profiles'' or simply as ``CD'', which is short for ``Complete Disagreement''.
\end{enumerate}

To reflect low to high levels of prior confidence in the efficacy data consistency across stages, we run the simulations per scenario with the prior probability of exchangeability $\omega = 0, 0.25, 0.5, 0.75, 1$. For scenarios under the alternative hypothesis $H_1$ we assume an effect size of 0.25 (i.e., in all stage II dose-efficacy profiles, the highest probability of efficacy is equal to $p_0 + 0.25 = 0.4$, with $p_0 = 0.15$). For scenarios under $H_0$, the highest probability of efficacy in stage II is equal to $p_0$. 

Overall, we have a large a number of comparisons given that for each dose-toxicity profile there are two different stage II dose-efficacy profiles, each coupled with three different stage I dose-efficacy profiles (i.e., CA, PA and CD). We have scenarios under $H_1$ and $H_0$, and furthermore five different values of $\omega$. To facilitate the communication of the simulation results over a large number of scenarios, we label the configuration of dose-toxicity and dose-efficacy profiles by Scenario A to H as follows:

\begin{itemize}
    \item Dose-toxicity profile 1 + stage II dose-efficacy profile 1 under $H_1$ = Scenario A,
    \item Dose-toxicity profile 1 + stage II dose-efficacy profile 2 under $H_1$ = Scenario B,
    \item Dose-toxicity profile 2 + stage II dose-efficacy profile 1 under $H_1$ = Scenario C,
    \item Dose-toxicity profile 2 + stage II dose-efficacy profile 2 under $H_1$ = Scenario D,
    \item Dose-toxicity profile 1 + stage II dose-efficacy profile 1 under $H_0$ = Scenario E,
    \item Dose-toxicity profile 1 + stage II dose-efficacy profile 2 under $H_0$ = Scenario F,
    \item Dose-toxicity profile 2 + stage II dose-efficacy profile 1 under $H_0$ = Scenario G,
    \item Dose-toxicity profile 2 + stage II dose-efficacy profile 2 under $H_0$ = Scenario H.
\end{itemize}

\begin{figure}[H]
\caption{True dose-efficacy profiles favoring the alternative hypothesis $H_1$ under each dose-toxicity scenarios varying with the dose of Cisplatin. In each efficacy scenario we have the true stage II efficacy profile (red), and three stage I efficacy scenarios: i) one that is exactly like the stage II dose-efficacy profile (in red), ii) one in which the optimal dose combination is the same but the efficacy profile is slightly different (green) and iii) one that is completely different to the stage II dose-efficacy profile (blue). The gray line represents the threshold $p_0 = 0.15$.}
\centering
\vspace{0.25cm}
\includegraphics[scale=0.7]{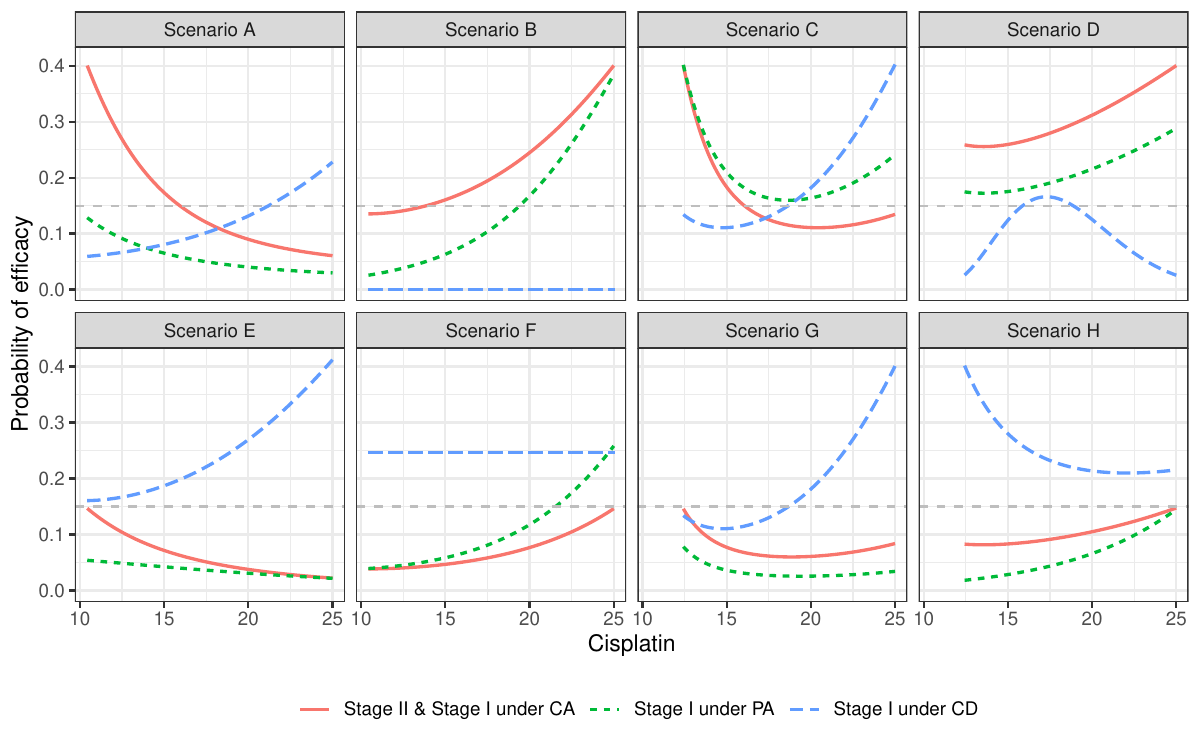}
\label{Figure_2}
\end{figure}

The true dose-efficacy profile per scenario, with specification of model parameters, is given in Table S3 in the supplementary material, and displayed graphically in Figure \ref{Figure_2}.

The sample sizes for Stages I and II are $N_1 = N_2 = 30$ and we simulated $J=1000$ trials using Algorithm \ref{algorithm1}. The DLT and efficacy responses were generated from models (\ref{modelprobdlt}) and (\ref{eq_dose_efficacy_model}), respectively.

\subsection{Results}
\label{sc_simulation_results}

As discussed in section \ref{sc_marginal_efficacy}, the value $\omega = 0$ implements no borrowing of information. In other words, the treatment efficacy in equation \eqref{eq_bvn} would be estimated using data from one stage solely, leading to a complete discard of stage I efficacy data. It is of interest to quantify the improvement achieved by allowing the combination of efficacy data based on the assumption of full exchangeability (with $\omega$ = 1) or partial exchangeability (with $0 < \omega < 1$).

\begin{figure}[H]
\caption{Differences in the probability of rejecting $H_0$ with respect to $\omega = 0$ in scenarios under the $H_1$ (i.e., power) and under the $H_0$ (i.e., type-I error). Scenario A-D and E-H correspond to settings under $H_1$ and $H_0$, respectively.}
\centering
\vspace{0.25cm}
\includegraphics[scale=0.7]{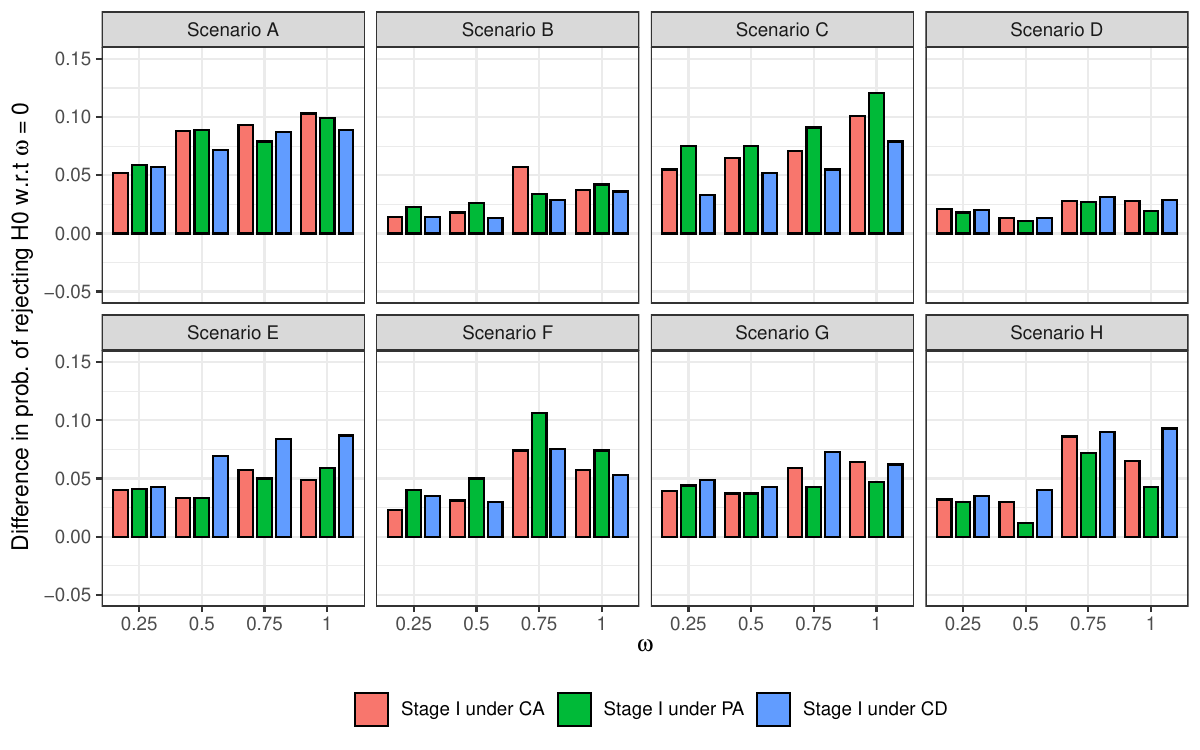}
\label{Figure_3}
\end{figure}

In Figure \ref{Figure_3}, we display the power and type-I error values obtained at different values of $\omega > 0$ with respect to $\omega = 0$. Under $H_0$ (i.e., scenarios E-H), if we choose low to medium values of $\omega$ (e.g., $0 < \omega \leq 0.25$), the type-I error varies between 0 and 0.049 with respect to $\omega = 0$, depending on the scenario and the level of agreement. That is, the type-I error remains very close to its reference value (i.e., with $\omega = 0$). With larger values of $\omega$ (i.e., $\omega > 0.25 $), the type-I error varies between 0.012 and 0.106 with respect to $\omega = 0$. Under $H_1$ (i.e., scenarios A-D), the power increases as the value of $\omega$, with differences, with respect to $\omega = 0$, up to 0.121. We notice that for values of $\omega \leq 0.25$ the power gain is already notable. The numerical results of power and type-I error for all values of $\omega$ are presented in the supplementary material (Figure S1). In scenarios under $H_1$ with $\omega = 0$, the power ranges between 0.66 and 0.93, whereas in scenarios under $H_0$ with $\omega = 0$, the type-I error ranges between 0.11 and 0.21. These power and type-I error results are consistent with those reported in previous publications \cite{tighiouart2019two,jimenez2020bayesian,jimenez2021combining}.

\begin{figure}[H]
\caption{Distribution of the recommended optimal dose combinations in scenarios under the alternative hypothesis (A-D) and levels of agreement CA, PA and CD. The black curve represents the MTD curve.}
\centering
\vspace{0.25cm}
\includegraphics[scale=0.6]{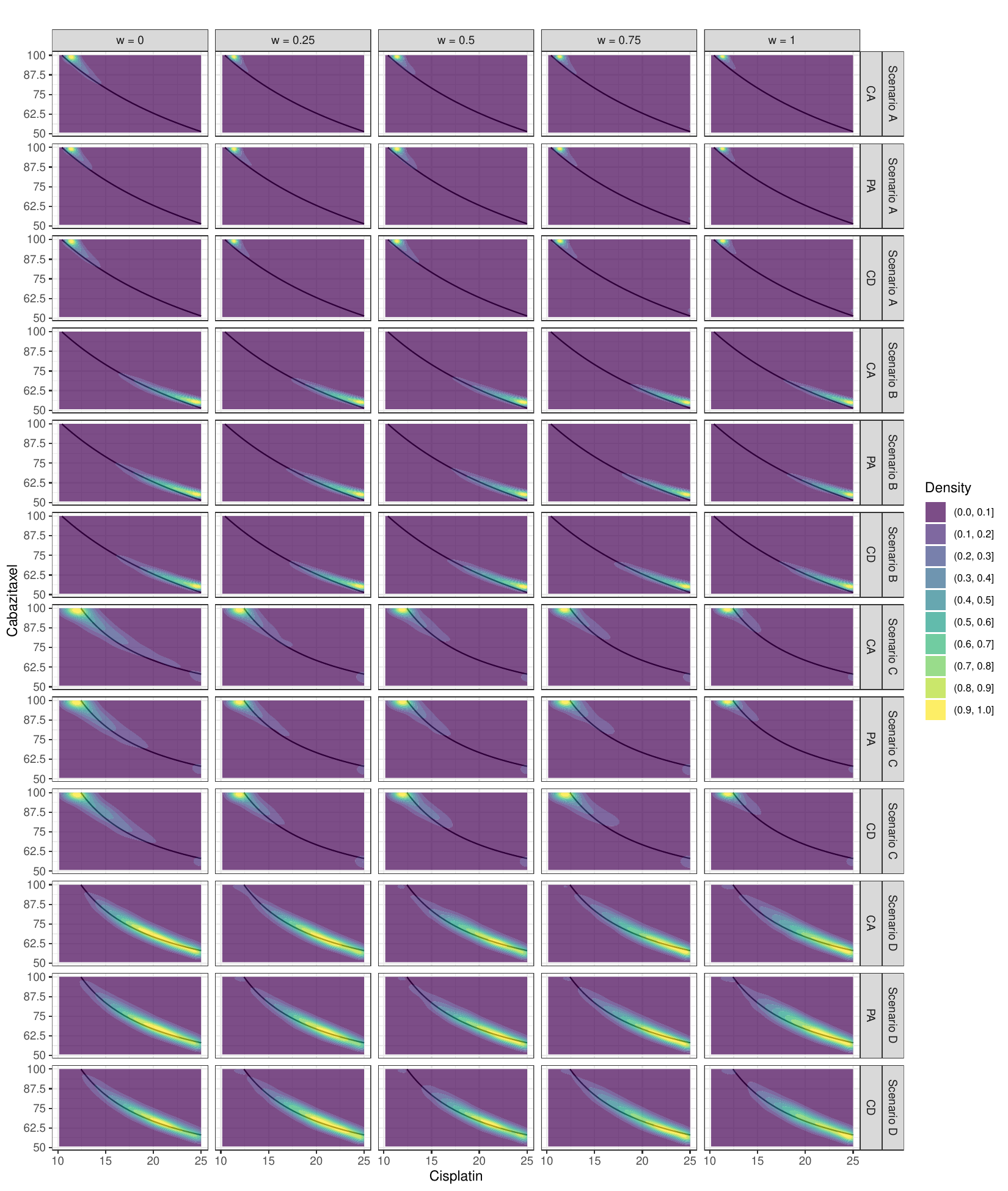}
\label{Figure_4}
\end{figure}

In Figure \ref{Figure_4} we present the distribution of the recommended optimal dose combinations across all scenarios, agreement levels and values of $\omega$. Overall, we see how the design correctly identifies the most efficacious region of the MTD curve by allocating there the majority of patients. At $\omega = 0$, we notice that the level of dispersion of the distribution is generally higher than with $\omega > 0$. As we increase the value of $\omega$, the dispersion shrinks towards the mode of the distribution. This behaviour is manifested under the levels of CA and PA. These results are as expected: our model effectively discounts efficacy data from stage I if it is not consistent with the efficacy data observed in stage II.

\begin{figure}[H]
\caption{Difference in the proportion of correct dose combination recommendation between models with $\omega > 0$ and $\omega = 0$ in settings under $H_1$.}
\centering
\vspace{0.25cm}
\includegraphics[scale=0.5]{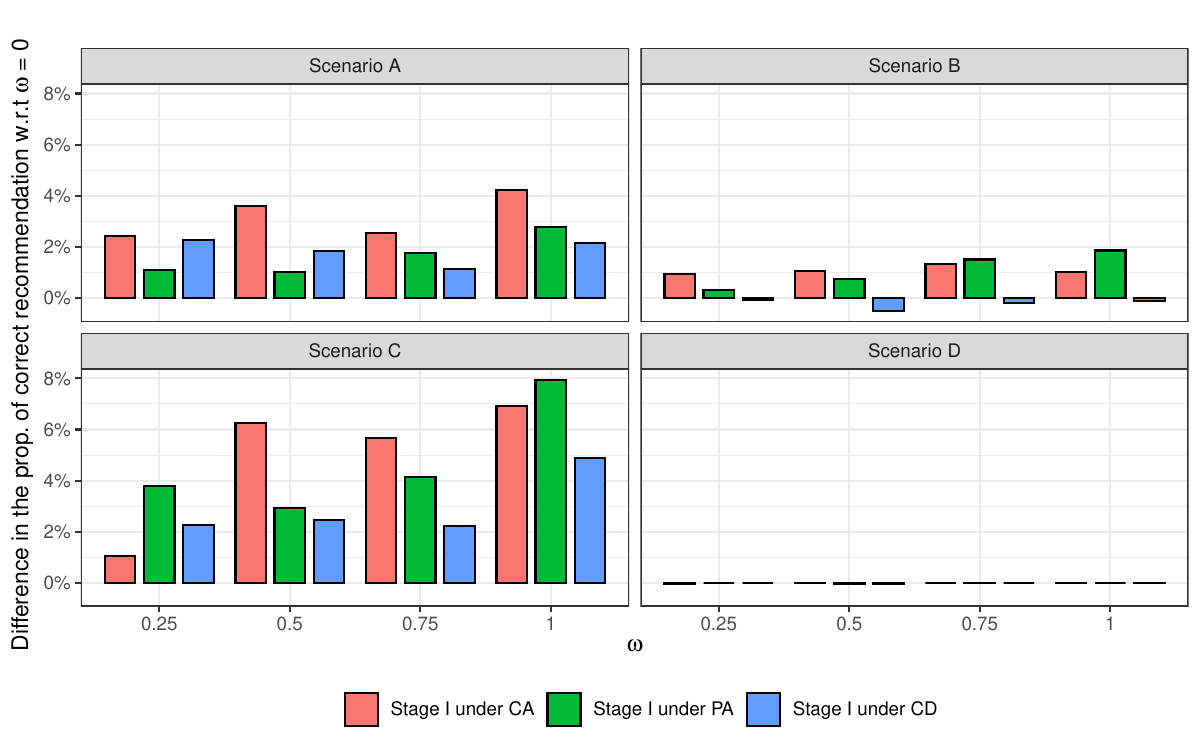}
\label{Figure_5}
\end{figure}

In Figure \ref{Figure_5}, we present the difference in the recommended optimal dose combinations with true probability of efficacy above $p_0$ (also known as the proportion of correct recommendation), between models with $\omega > 0$ and $\omega = 0$. In settings with $\omega = 0$, the proportion of correct recommendation ranges between 81-100\%, which is consistent with the values reported in previous publications\cite{tighiouart2019two,jimenez2020bayesian,jimenez2021combining}. For values of $\omega > 0$, such proportion varies between -0.49\% and 7.95\%. In scenario D, the proportion of correct recommendation with $\omega = 0$ is already practically 100\%, which remains the same for values of $\omega > 0$. This explains why the difference in the proportion of correct recommendation between $\omega = 0$ and $\omega > 0$ is approximately 0.

\begin{figure}[t]
\caption{Difference in the proportion of trials with early stopping for futility, between models with $\omega > 0$ and $\omega = 0$ under $H_1$ (A-D) and $H_0$ (E-H).}
\centering
\vspace{0.25cm}
\includegraphics[scale=0.7]{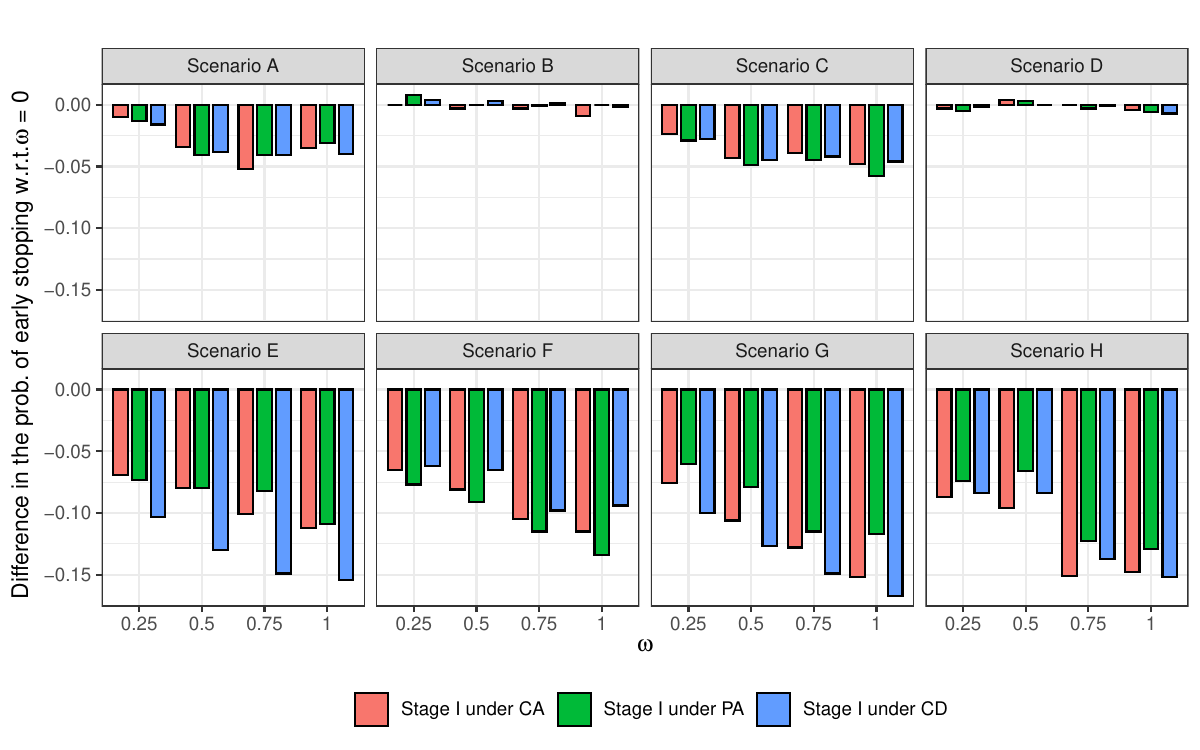}
\label{Figure_6}
\end{figure}

In Figure \ref{Figure_6}, we show the differences in terms of the probability of early stopping for futility, under both $H_1$ and $H_0$. At $\omega = 0$, the probability of early stopping for futility ranges between 0.012 and 0.126 under $H_1$, and between 0.478 and 0.612 under $H_0$, depending on the scenario. Under $H_1$, by increasing $\omega$, we see that the probability of early stopping for futility varies between -0.058 and 0.008, with respect to $\omega = 0$. Under $H_0$, by increasing $\omega$, we see that the probability of early stopping for futility varies between -0.167 and -0.060, with respect to $\omega = 0$. It is worth mentioning that in scenarios under $H_1$, the probabilities of early stopping for futility with $\omega = 0$ are already low, and therefore it is reasonable that allowing for robust sharing of efficacy data across stages does not have a major impact on the probability of early stopping for futility. On the other hand, in scenarios under $H_0$, the probability of early stopping for futility with $\omega = 0$ is, as expected, higher and a big decrease would be problematic. However, we see that by selecting a conservative value of $\omega$, such as $\omega = 0.25$, the decrease in the probability of early stopping is usually lower than 0.1.

In Table S5 of the supplementary material, we show the average sample sizes obtained when applying this early stopping rule. Under $H_1$, results show that the observed decrease in the probability of early stopping for futility caused by the increment of $\omega$ translated into an average sample size increase of 0-1 patients with respect to $\omega = 0$. Under $H_0$, the increment of $\omega$ translated into an average sample size increase of 0-2 patients with respect to $\omega = 0$.

In Figure S2 of the supplementary material, we present the difference in the proportion of patients allocated to dose combinations with true probability of efficacy above $p_0$ in stage II, between models with $\omega > 0$ and $\omega = 0$.  With $\omega = 0$, the proportion of patients allocated to dose combinations with true probability of efficacy above $p_0$ in stage II ranges between 40-95\%, which is consistent with the values reported by \cite{tighiouart2019two,jimenez2020bayesian,jimenez2021combining}. As we increase the value of $\omega$ this proportion increases between 0.25\% and 6.17\%. 

In terms of safety, we observe that scenarios A, B, E and F (i.e., dose-toxicity scenario 1 in Figure \ref{Figure_1}) have an overall (i.e., stage I + stage II) average DLT rate between 27\% and 35\%, depending on the scenario, with an average proportion of trials with DLT rate above $\theta_T + 0.1$ of 0\%. Stage II alone in these scenarios has an average DLT rate between 28\% and 43\%, with an average proportion of trials with DLT rate above $\theta_T + 0.1$ of 7\% and 42\%. Scenarios C, D, G and H (i.e., dose-toxicity scenario 2 in Figure \ref{Figure_1}) have an overall (i.e., stage I + stage II) average DLT rate between 26\% and 35\%, depending on the scenario, with an average proportion of trials with DLT rate above $\theta_T + 0.1$ of 0\%. Stage II alone in these scenarios has an average DLT rate between 28\% and 43\%, with an average proportion of trials with DLT rate above $\theta_T + 0.1$ between 7\% and 43\%. Because toxicity data is not shared across stages, these values are constant across all values of $\omega$. In Figure S3 of the supplementary material, we display the probability of early stopping for safety. In scenarios A, B, E and F, this probability is close 0.25 whereas in scenarios C, D, G and H is close to 0.05. These values are consistent with the informative prior distributions for the dose-toxicity models and the distance between the dose combination 15/75 $mg/m^2$, which has a prior probability if DLT of approximately 0.33, and the true MTD curves. Also, because toxicity data is not shared across stages, we do not present the average sample sizes as they are in line with those reported by \cite{jimenez2020bayesian} in a similar setting.

\section{Discussions}
\label{sc_discussion}

Motivated by a real phase I-II trial that combines continuous dose levels of cisplatin and cabazitaxel involving two different populations of patients with advanced prostate cancer, in this paper we present a phase I-II design in two stages that allows robust integration of efficacy data across relevant patient populations. The main contribution of this article lies in the formal consideration about the uncertainty around the potentially different dose-efficacy profiles across stages.  We propose to employ a robust Bayesian hierarchical random-effects model to allow sharing of information on the efficacy across stages, assuming that the related parameters are either exchangeable or nonexchangeable. In other words, the key idea is to exploit any potential similarities between the dose-efficacy profiles so as to borrow information, while avoiding too optimistic borrowing under the presence of data inconsistency across stages. This proposal requires specification of the prior probability that the main effects set of parameters are exchangeable across stages. We denote this prior probability by $\omega$ which in practice is selected by subject-matter experts. When $\omega = 0$, the design estimates the stage II dose-efficacy profile independently from stage I \cite{tighiouart2019two} and \cite{jimenez2020bayesian}. In this article, we focus on analyzing the operating characteristics of stage II, and we study how these vary, with respect to $\omega = 0$, as we increase the prior probability of exchangeability $\omega$ under different stage I dose-efficacy profiles. For a detailed evaluation of the stage I operating characteristics we refer the reader to Tighiouart (2019)\cite{tighiouart2019two}, Jim\'enez et al. (2020)\cite{jimenez2020bayesian} and Jim\'enez and Tighiouart (2022) \cite{jimenez2021combining}.

The selection of the dose-efficacy data model is closely related to the type of compound investigated in a phase I-II clinical trial. With cytotoxic agents the monotonicity assumption is expected to hold also from an efficacy perspective (i.e., a compound will have greater activity as the dose increases). Thus a linear model such as the one defined in \eqref{eq_dose_efficacy_model} will be sufficient to capture the dose-efficacy relationship. However, with other types of compounds such as molecularly targeted therapies, more flexible modelling approaches may be needed to capture dose-efficacy relationships where the probability of efficacy may not even increase with the dose.

We have limited the simulations to the two main dose-toxicity scenarios considered by the principal investigator of the motivating example. In each of these dose-toxicity profiles, we have studied two different stage II dose-efficacy profiles, each one accompanied with three stage I dose-efficacy profiles that have different levels of similarity with respect to the stage II dose-efficacy profile. Also, because we allow the main effects set of parameters to be exchangeable across stages, similarity or agreement across stages is based only on these two parameters. However, depending on the application and the definition of the dose-efficacy profile, this work could be extended by tweaking the \texttt{JAGS} code, which we have made publicly available, so it includes other parameters in the set of parameters that could be exchangeable across stages.

The evaluation we present in this article aims to assess whether the overall operating characteristics of design improve by allowing robust integration of efficacy data across stages in scenarios under complete agreement, partial agreement and complete disagreement between the stage I and stage II dose-efficacy profiles. In other words, we aim to evaluate how much we can benefit from sharing efficacy data across stages when efficacy the data is completely or partially consistent across stages based on the main effects set of parameters, but also to what extend we expect to penalize the design's operating characteristics when the efficacy data is inconsistent across stages. The assessment is done in the original setting with continuous dose combination levels under $H_0$ and $H_1$ following the case study described in section \ref{sc_motivating_example}. 

In scenarios favoring the alternative hypothesis and visualized in Figure \ref{Figure_2} we observe a generalized improvement of the operating characteristics by permitting sharing the efficacy data across stages (i.e., $\omega > 0$). The degree of improvement would depend however on the genuine extent of consistency between the stage-wise efficacy profiles and of course on the value of $\omega$ that we select. We note that with $\omega = 0.25$ there is already a considerable improvement of the designs operating characteristics in comparison with higher values of $\omega$. 

In scenarios under the null hypothesis we observed a small inflation in the type-I error and a slight decrease in the probability of early stopping for futility. Under this hypothesis, having a high value of $\omega$ in situations of complete disagreement across the stage I and stage II efficacy profiles generally yields the worst performance. However, with $\omega = 0.25$, the differences are much smaller with respect to settings in which there is complete or partial agreement across the stage I and stage II efficacy profiles. 

Overall, we believe that allowing for sharing of efficacy data across stages increases the probability of finding an appropriate dose combination for further phase III studies. However, this approach requires preliminary knowledge on the drug combination. We regard this as acceptable, since there is not a unique design configuration that will fit all applications. For example, in our proposal we allow the main effects set of parameters to be exchangeable or nonexchangeable across stages, and thus our definition of similarity or agreement across stages is based solely on the main effects parameters. Moreover, we have seen that two dose-efficacy profiles that are similar in terms of the two main effects set of parameters can have completely different intercepts, and can potentially induce either a power loss or a type-I error inflation. Therefore, a clear understanding of what is considered ``similar'' is key to decide how we want to synthesise the efficacy data across stages. A potential solution to this problem could be to include the intercept in the set of parameters for the assumption of exchangeability or non-exchangeability.

One potential extension of the methodology presented in this manuscript, which we plan to explore in the future, is to robustly combine toxicity data across stages in this particular setting. By doing so, we would eliminate the assumption that the dose-toxicity profiles are equivalent across different patient population and we would account for population-specific characteristics with respect to the MTD.

\subsection*{Data availability}

The \texttt{R} and \texttt{JAGS} scripts needed to fully reproduce the results presented in this article are available at \newline
\href{https://github.com/jjimenezm1989/Bayesian-phase-I-II-design-combining-efficacy-data}{\texttt{https://github.com/jjimenezm1989/Bayesian-phase-I-II-design-combining-efficacy-data}}

\subsection*{Funding}

Dr. Zheng's contribution to this manuscript was supported by Cancer Research UK (RCCPDF$\backslash$100008).

\subsection*{Acknowledgements}

We thank the associate editor, two referees, and the editor for the constructive comments that led to significant improvements in the article.
\newline
\newline
Dr. Jim\'enez is employed by Novartis Pharma AG, who provided support in the form of salary for the author but did not have any additional role in the preparation of the manuscript. Also, the views expressed in this publication are those of the authors and should not be attributed to any of the funding institutions or organisations to which the authors are affiliated.

\bibliographystyle{plainnat}
\bibliography{references.bib}

\begin{thebibliography}{28}
\providecommand{\natexlab}[1]{#1}
\providecommand{\url}[1]{\texttt{#1}}
\expandafter\ifx\csname urlstyle\endcsname\relax
  \providecommand{\doi}[1]{doi: #1}\else
  \providecommand{\doi}{doi: \begingroup \urlstyle{rm}\Url}\fi

\bibitem[Babb et~al.(1998)Babb, Rogatko, and Zacks]{babb1998cancer}
James Babb, Andr{\'e} Rogatko, and Shelemyahu Zacks.
\newblock Cancer phase i clinical trials: efficient dose escalation with
  overdose control.
\newblock \emph{Statistics in medicine}, 17\penalty0 (10):\penalty0 1103--1120,
  1998.

\bibitem[Cai et~al.(2014)Cai, Yuan, and Ji]{cai2014bayesian}
Chunyan Cai, Ying Yuan, and Yuan Ji.
\newblock A bayesian dose-finding design for oncology clinical trials of
  combinational biological agents.
\newblock \emph{Journal of the Royal Statistical Society. Series C, Applied
  statistics}, 63\penalty0 (1):\penalty0 159, 2014.

\bibitem[Gasparini(2013)]{gasparini2013general}
Mauro Gasparini.
\newblock General classes of multiple binary regression models in dose finding
  problems for combination therapies.
\newblock \emph{Journal of the Royal Statistical Society: Series C (Applied
  Statistics)}, 62\penalty0 (1):\penalty0 115--133, 2013.

\bibitem[Ivanova(2003)]{ivanova2003new}
Anastasia Ivanova.
\newblock A new dose-finding design for bivariate outcomes.
\newblock \emph{Biometrics}, 59\penalty0 (4):\penalty0 1001--1007, 2003.

\bibitem[Ivanova et~al.(2009)Ivanova, Liu, Snyder, and
  Snavely]{ivanova2009adaptive}
Anastasia Ivanova, Ken Liu, Ellen Snyder, and Duane Snavely.
\newblock An adaptive design for identifying the dose with the best
  efficacy/tolerability profile with application to a crossover dose-finding
  study.
\newblock \emph{Statistics in medicine}, 28\penalty0 (24):\penalty0 2941--2951,
  2009.

\bibitem[Jiang et~al.(2021)Jiang, Li, Yan, Yap, and Yuan]{jiang2021shotgun}
Liyun Jiang, Ruobing Li, Fangrong Yan, Timothy~A Yap, and Ying Yuan.
\newblock Shotgun: A bayesian seamless phase i-ii design to accelerate the
  development of targeted therapies and immunotherapy.
\newblock \emph{Contemporary Clinical Trials}, 104:\penalty0 106338, 2021.

\bibitem[Jim\'enez and Tighiouart(2022)]{jimenez2021combining}
Jos\'e~L. Jim\'enez and Mourad Tighiouart.
\newblock Combining cytotoxic agents with continuous dose levels in seamless
  phase i-ii clinical trials.
\newblock \emph{Journal of the Royal Statistical Society: Series C (Applied
  Statistics)}, 71\penalty0 (5):\penalty0 1996--2013, 2022.

\bibitem[Jimenez et~al.(2019)Jimenez, Tighiouart, and
  Gasparini]{jimenez2019cancer}
Jose~L. Jimenez, Mourad Tighiouart, and Mauro Gasparini.
\newblock Cancer phase i trial design using drug combinations when a fraction
  of dose limiting toxicities is attributable to one or more agents.
\newblock \emph{Biometrical Journal}, 61\penalty0 (2):\penalty0 319--332, 2019.

\bibitem[Jim{\'e}nez et~al.(2020)Jim{\'e}nez, Kim, and
  Tighiouart]{jimenez2020bayesian}
Jos{\'e}~L Jim{\'e}nez, Sungjin Kim, and Mourad Tighiouart.
\newblock A bayesian seamless phase i--ii trial design with two stages for
  cancer clinical trials with drug combinations.
\newblock \emph{Biometrical Journal}, 62\penalty0 (5):\penalty0 1300--1314,
  2020.

\bibitem[Le~Tourneau et~al.(2009)Le~Tourneau, Lee, and Siu]{le2009dose}
Christophe Le~Tourneau, J~Jack Lee, and Lillian~L Siu.
\newblock Dose escalation methods in phase i cancer clinical trials.
\newblock \emph{JNCI: Journal of the National Cancer Institute}, 101\penalty0
  (10):\penalty0 708--720, 2009.

\bibitem[Lin et~al.(2020)Lin, Thall, and Yuan]{lin2020adaptive}
Ruitao Lin, Peter~F Thall, and Ying Yuan.
\newblock An adaptive trial design to optimize dose-schedule regimes with
  delayed outcomes.
\newblock \emph{Biometrics}, 76\penalty0 (1):\penalty0 304--315, 2020.

\bibitem[Liu et~al.(2018)Liu, Guo, and Yuan]{liu2018bayesian}
Suyu Liu, Beibei Guo, and Ying Yuan.
\newblock A bayesian phase i/ii trial design for immunotherapy.
\newblock \emph{Journal of the American Statistical Association}, 113\penalty0
  (523):\penalty0 1016--1027, 2018.

\bibitem[Lockhart et~al.(2014)Lockhart, Sundaram, Sarantopoulos, Mita,
  Wang-Gillam, Moseley, Barber, Lane, Wack, Kassalow,
  et~al.]{lockhart2014phase}
A~Craig Lockhart, Shankar Sundaram, John Sarantopoulos, Monica~M Mita, Andrea
  Wang-Gillam, Jennifer~L Moseley, Stephanie~L Barber, Alex~R Lane, Claudine
  Wack, Laurent Kassalow, et~al.
\newblock Phase i dose-escalation study of cabazitaxel administered in
  combination with cisplatin in patients with advanced solid tumors.
\newblock \emph{Investigational new drugs}, 32\penalty0 (6):\penalty0
  1236--1245, 2014.

\bibitem[Lyu et~al.(2019)Lyu, Ji, Zhao, and Catenacci]{lyu2019aaa}
Jiaying Lyu, Yuan Ji, Naiqing Zhao, and Daniel~VT Catenacci.
\newblock Aaa: triple adaptive bayesian designs for the identification of
  optimal dose combinations in dual-agent dose finding trials.
\newblock \emph{Journal of the Royal Statistical Society. Series C, Applied
  statistics}, 68\penalty0 (2):\penalty0 385, 2019.

\bibitem[Neuenschwander et~al.(2016)Neuenschwander, Roychoudhury, and
  Schmidli]{neuenschwander2016use}
Beat Neuenschwander, Satrajit Roychoudhury, and Heinz Schmidli.
\newblock On the use of co-data in clinical trials.
\newblock \emph{Statistics in Biopharmaceutical Research}, 8\penalty0
  (3):\penalty0 345--354, 2016.

\bibitem[Rogatko et~al.(2008)Rogatko, Ghosh, Vidakovic, and
  Tighiouart]{rogatko2008patient}
Andr{\'e} Rogatko, Pulak Ghosh, Brani Vidakovic, and Mourad Tighiouart.
\newblock Patient-specific dose adjustment in the cancer clinical trial
  setting.
\newblock \emph{Pharmaceutical Medicine}, 22\penalty0 (6):\penalty0 345--350,
  2008.

\bibitem[Roychoudhury and Neuenschwander(2020)]{roychoudhury2020bayesian}
Satrajit Roychoudhury and Beat Neuenschwander.
\newblock Bayesian leveraging of historical control data for a clinical trial
  with time-to-event endpoint.
\newblock \emph{Statistics in medicine}, 39\penalty0 (7):\penalty0 984--995,
  2020.

\bibitem[Shi and Yin(2013)]{shi2013escalation}
Yun Shi and Guosheng Yin.
\newblock Escalation with overdose control for phase i drug-combination trials.
\newblock \emph{Statistics in medicine}, 32\penalty0 (25):\penalty0 4400--4412,
  2013.

\bibitem[Thall and Cook(2004)]{thall2004dose}
Peter~F Thall and John~D Cook.
\newblock Dose-finding based on efficacy-toxicity trade-offs.
\newblock \emph{Biometrics}, 60\penalty0 (3):\penalty0 684--693, 2004.

\bibitem[Tighiouart(2019)]{tighiouart2019two}
Mourad Tighiouart.
\newblock Two-stage design for phase i--ii cancer clinical trials using
  continuous dose combinations of cytotoxic agents.
\newblock \emph{Journal of the Royal Statistical Society: Series C (Applied
  Statistics)}, 68\penalty0 (1):\penalty0 235--250, 2019.

\bibitem[Tighiouart and Rogatko(2012)]{tighiouart2012number}
Mourad Tighiouart and Andre Rogatko.
\newblock Number of patients per cohort and sample size considerations using
  dose escalation with overdose control.
\newblock \emph{Journal of Probability and Statistics}, 2012, 2012.

\bibitem[Tighiouart et~al.(2005)Tighiouart, Rogatko, and
  Babb]{tighiouart2005flexible}
Mourad Tighiouart, Andr{\'e} Rogatko, and James~S Babb.
\newblock Flexible bayesian methods for cancer phase i clinical trials. dose
  escalation with overdose control.
\newblock \emph{Statistics in medicine}, 24\penalty0 (14):\penalty0 2183--2196,
  2005.

\bibitem[Tighiouart et~al.(2010)Tighiouart, Rogatko,
  et~al.]{tighiouart2010dose}
Mourad Tighiouart, Andr{\'e} Rogatko, et~al.
\newblock Dose finding with escalation with overdose control (ewoc) in cancer
  clinical trials.
\newblock \emph{Statistical Science}, 25\penalty0 (2):\penalty0 217--226, 2010.

\bibitem[Tighiouart et~al.(2017)Tighiouart, Li, and
  Rogatko]{tighiouart2017bayesian}
Mourad Tighiouart, Quanlin Li, and Andr{\'e} Rogatko.
\newblock A bayesian adaptive design for estimating the maximum tolerated dose
  curve using drug combinations in cancer phase i clinical trials.
\newblock \emph{Statistics in medicine}, 36\penalty0 (2):\penalty0 280--290,
  2017.

\bibitem[Wheeler et~al.(2017)Wheeler, Sweeting, and
  Mander]{wheeler2017toxicity}
Graham~M Wheeler, Michael~J Sweeting, and Adrian~P Mander.
\newblock Toxicity-dependent feasibility bounds for the escalation with
  overdose control approach in phase i cancer trials.
\newblock \emph{Statistics in medicine}, 36\penalty0 (16):\penalty0 2499--2513,
  2017.

\bibitem[Yuan and Yin(2009)]{yuan2009bayesian}
Ying Yuan and Guosheng Yin.
\newblock Bayesian dose finding by jointly modelling toxicity and efficacy as
  time-to-event outcomes.
\newblock \emph{Journal of the Royal Statistical Society: Series C (Applied
  Statistics)}, 58\penalty0 (5):\penalty0 719--736, 2009.

\bibitem[Zheng et~al.(2020)Zheng, Hampson, and Wandel]{zheng2020robust}
Haiyan Zheng, Lisa~V Hampson, and Simon Wandel.
\newblock A robust bayesian meta-analytic approach to incorporate animal data
  into phase i oncology trials.
\newblock \emph{Statistical methods in medical research}, 29\penalty0
  (1):\penalty0 94--110, 2020.

\bibitem[Zheng et~al.(2021)Zheng, Hampson, and Jaki]{zheng2021bridging}
Haiyan Zheng, Lisa~V Hampson, and Thomas Jaki.
\newblock Bridging across patient subgroups in phase i oncology trials that
  incorporate animal data.
\newblock \emph{Statistical Methods in Medical Research}, 30\penalty0
  (4):\penalty0 1057--1071, 2021.

\end{thebibliography}

\newpage

\section*{Supplementary Material}

\newcommand{\hbAppendixPrefix}{S}
\renewcommand{\thefigure}{\hbAppendixPrefix\arabic{figure}}
\setcounter{figure}{0}
\renewcommand{\thetable}{\hbAppendixPrefix\arabic{table}} 
\setcounter{table}{0}
\renewcommand{\theequation}{\hbAppendixPrefix\arabic{equation}} 
\setcounter{equation}{0}

\begin{table}[h]
\caption{Median prior probability of efficacy with 95\% credible intervals using weakly informative prior distributions}
\centering
\vspace{0.25cm}
\begin{tabular}{|c|c|c|c|c|}
\hline
\multirow{2}{*}{Cisplatin ($mg/m^2$)} & \multirow{2}{*}{Cabazitaxel ($mg/m^2$)} & \multicolumn{3}{c|}{Median (95\% Credible Interval)} \\ \cline{3-5} 
 &  & $w = 0$ & $w = 0.5$ & $w = 1$ \\ \hline
10 & 50 & 0.14 (0 - 0.99) & 0.14 (0 - 0.99) & 0.14 (0 - 0.99) \\ \hline
10 & 100 & 0.41 (0 - 1) & 0.41 (0 - 1) & 0.42 (0 - 1) \\ \hline
25  & 50  & 0.42 (0 - 1) & 0.42 (0 - 1) & 0.42 (0 - 1) \\ \hline
25 & 100 & 0.84 (0 - 1) & 0.85 (0 - 1) & 0.86 (0 - 1) \\ \hline
17.5 &75 & 0.48 (0 - 1) & 0.5 (0 - 1) & 0.5 (0 - 1) \\ \hline
15 & 75 & 0.41 (0 - 1) & 0.41 (0 - 1) & 0.41 (0 - 1) \\ \hline
\end{tabular}
\label{table_prior_efficacy}
\end{table}

\begin{table}[h]
\caption{True dose-toxicity model parameter values of the 2 scenarios taken from Tighiouart (2019).}
\centering
\vspace{0.25cm}
\begin{tabular}{|c|c|c|}
\hline
\multicolumn{3}{|c|}{\makecell{Stage I \& II  dose-toxicity \\ profile parameters}} \\ \hline
 & Profile 1 & Profile 2 \\ \hline
$\rho_{00}$  & $1 \times 10^{-7}$ & 0.001 \\ \hline
$\rho_{01}$  & 0.2 & 0.05 \\ \hline
$\rho_{10}$  & 0.2 & 0.05 \\ \hline
$\alpha_3$ & 10 & 10 \\ \hline
\end{tabular}
\label{table_toxicity_scenarios}
\end{table}

We recall that, to facilitate the comprehension of the simulation study, we rename the dose-toxicity and dose-efficacy scenarios as follows:

\begin{itemize}
    \item Dose-toxicity profile 1 + stage II dose-efficacy profile 1 under $H_1$ = Scenario A,
    \item Dose-toxicity profile 1 + stage II dose-efficacy profile 2 under $H_1$ = Scenario B,
    \item Dose-toxicity profile 2 + stage II dose-efficacy profile 1 under $H_1$ = Scenario C,
    \item Dose-toxicity profile 2 + stage II dose-efficacy profile 2 under $H_1$ = Scenario D,
    \item Dose-toxicity profile 1 + stage II dose-efficacy profile 1 under $H_0$ = Scenario E,
    \item Dose-toxicity profile 1 + stage II dose-efficacy profile 2 under $H_0$ = Scenario F,
    \item Dose-toxicity profile 2 + stage II dose-efficacy profile 1 under $H_0$ = Scenario G,
    \item Dose-toxicity profile 2 + stage II dose-efficacy profile 2 under $H_0$ = Scenario H.
\end{itemize}

\begin{table}[h]
\centering
\caption{True dose-efficacy model parameter values for stage I and II scenarios under $H_1$.}
\vspace{0.25cm}
\begin{tabular}{|c|ccccc|cccc|}
\hline
\multirow{2}{*}{}           & \multicolumn{5}{c|}{Stage I dose-efficacy paramters}                                                                                                            & \multicolumn{4}{c|}{Stage II dose-efficacy parameters}                                                                                                       \\ \cline{2-10} 
                            & \multicolumn{1}{c|}{\makecell{Agreement \\ level}} & \multicolumn{1}{c|}{$\beta_{01}$} & \multicolumn{1}{c|}{$\beta_{11}$} & \multicolumn{1}{c|}{$\beta_{21}$} & $\beta_{23}$ & \multicolumn{1}{c|}{$\beta_{02}$}        & \multicolumn{1}{c|}{$\beta_{12}$}            & \multicolumn{1}{c|}{$\beta_{22}$}           & $\beta_{32}$         \\ \hline
\multirow{3}{*}{Scenario A} & \multicolumn{1}{c|}{CA}              & \multicolumn{1}{c|}{-5}           & \multicolumn{1}{c|}{0.75}         & \multicolumn{1}{c|}{1.51}         & 0.5          & \multicolumn{1}{c|}{\multirow{3}{*}{-5}} & \multicolumn{1}{c|}{\multirow{3}{*}{0.75}}   & \multicolumn{1}{c|}{\multirow{3}{*}{1.51}}  & \multirow{3}{*}{0.5} \\ \cline{2-6}
                            & \multicolumn{1}{c|}{PA}              & \multicolumn{1}{c|}{-5}           & \multicolumn{1}{c|}{0.35}         & \multicolumn{1}{c|}{1.11}         & 0.5          & \multicolumn{1}{c|}{}                    & \multicolumn{1}{c|}{}                        & \multicolumn{1}{c|}{}                       &                      \\ \cline{2-6}
                            & \multicolumn{1}{c|}{CD}              & \multicolumn{1}{c|}{-5}           & \multicolumn{1}{c|}{1.31}         & \multicolumn{1}{c|}{0.75}         & 0.5          & \multicolumn{1}{c|}{}                    & \multicolumn{1}{c|}{}                        & \multicolumn{1}{c|}{}                       &                      \\ \hline
\multirow{3}{*}{Scenario B} & \multicolumn{1}{c|}{CA}              & \multicolumn{1}{c|}{-5}           & \multicolumn{1}{c|}{1.5035}       & \multicolumn{1}{c|}{1.1}          & 0.5          & \multicolumn{1}{c|}{\multirow{3}{*}{-5}} & \multicolumn{1}{c|}{\multirow{3}{*}{1.5035}} & \multicolumn{1}{c|}{\multirow{3}{*}{1.1}}   & \multirow{3}{*}{0.5} \\ \cline{2-6}
                            & \multicolumn{1}{c|}{PA}              & \multicolumn{1}{c|}{-5}           & \multicolumn{1}{c|}{1.5}          & \multicolumn{1}{c|}{0.2}          & 0.5          & \multicolumn{1}{c|}{}                    & \multicolumn{1}{c|}{}                        & \multicolumn{1}{c|}{}                       &                      \\ \cline{2-6}
                            & \multicolumn{1}{c|}{CD}              & \multicolumn{1}{c|}{-8}           & \multicolumn{1}{c|}{-10}          & \multicolumn{1}{c|}{-10}          & 0            & \multicolumn{1}{c|}{}                    & \multicolumn{1}{c|}{}                        & \multicolumn{1}{c|}{}                       &                      \\ \hline
\multirow{3}{*}{Scenario C} & \multicolumn{1}{c|}{CA}              & \multicolumn{1}{c|}{-6}           & \multicolumn{1}{c|}{1.2}          & \multicolumn{1}{c|}{1.623}        & 0            & \multicolumn{1}{c|}{\multirow{3}{*}{-6}} & \multicolumn{1}{c|}{\multirow{3}{*}{1.2}}    & \multicolumn{1}{c|}{\multirow{3}{*}{1.623}} & \multirow{3}{*}{0}   \\ \cline{2-6}
                            & \multicolumn{1}{c|}{PA}              & \multicolumn{1}{c|}{-6}           & \multicolumn{1}{c|}{1.4}          & \multicolumn{1}{c|}{1.6}          & 0            & \multicolumn{1}{c|}{}                    & \multicolumn{1}{c|}{}                        & \multicolumn{1}{c|}{}                       &                      \\ \cline{2-6}
                            & \multicolumn{1}{c|}{CD}              & \multicolumn{1}{c|}{-6}           & \multicolumn{1}{c|}{1.623}        & \multicolumn{1}{c|}{1.2}          & 0            & \multicolumn{1}{c|}{}                    & \multicolumn{1}{c|}{}                        & \multicolumn{1}{c|}{}                       &                      \\ \hline
\multirow{3}{*}{Scenario D} & \multicolumn{1}{c|}{CA}              & \multicolumn{1}{c|}{-4}           & \multicolumn{1}{c|}{1.025}        & \multicolumn{1}{c|}{0.7}          & 3            & \multicolumn{1}{c|}{\multirow{3}{*}{-4}} & \multicolumn{1}{c|}{\multirow{3}{*}{1.025}}  & \multicolumn{1}{c|}{\multirow{3}{*}{0.7}}   & \multirow{3}{*}{3}   \\ \cline{2-6}
                            & \multicolumn{1}{c|}{PA}              & \multicolumn{1}{c|}{-4.5}         & \multicolumn{1}{c|}{1.025}        & \multicolumn{1}{c|}{0.7}          & 3            & \multicolumn{1}{c|}{}                    & \multicolumn{1}{c|}{}                        & \multicolumn{1}{c|}{}                       &                      \\ \cline{2-6}
                            & \multicolumn{1}{c|}{CD}              & \multicolumn{1}{c|}{-8}           & \multicolumn{1}{c|}{-5}           & \multicolumn{1}{c|}{-5}           & 27           & \multicolumn{1}{c|}{}                    & \multicolumn{1}{c|}{}                        & \multicolumn{1}{c|}{}                       &                      \\ \hline
\end{tabular}
\label{table_efficacy_scenarios}
\end{table}

\begin{table}[h]
\centering
\caption{True dose-efficacy model parameter values for stage I and II scenarios under $H_0$.}
\vspace{0.25cm}
\begin{tabular}{|c|ccccc|cccc|}
\hline
\multirow{2}{*}{} & \multicolumn{5}{c|}{Stage I dose-efficacy paramters}                                                                                                            & \multicolumn{4}{c|}{Stage II dose-efficacy parameters}                                                                                                          \\ \cline{2-10} 
                            & \multicolumn{1}{c|}{\makecell{Agreement \\ level}} & \multicolumn{1}{c|}{$\beta_{01}$} & \multicolumn{1}{c|}{$\beta_{11}$} & \multicolumn{1}{c|}{$\beta_{21}$} & $\beta_{23}$ & \multicolumn{1}{c|}{$\beta_{02}$}           & \multicolumn{1}{c|}{$\beta_{12}$}            & \multicolumn{1}{c|}{$\beta_{22}$}           & $\beta_{32}$         \\ \hline
\multirow{3}{*}{Scenario E} & \multicolumn{1}{c|}{CA}              & \multicolumn{1}{c|}{-4}           & \multicolumn{1}{c|}{-2}           & \multicolumn{1}{c|}{0.8}          & 0.5          & \multicolumn{1}{c|}{\multirow{3}{*}{-4}}    & \multicolumn{1}{c|}{\multirow{3}{*}{-2}}     & \multicolumn{1}{c|}{\multirow{3}{*}{0.8}}   & \multirow{3}{*}{0.5} \\ \cline{2-6}
                            & \multicolumn{1}{c|}{PA}              & \multicolumn{1}{c|}{-4}           & \multicolumn{1}{c|}{-2}           & \multicolumn{1}{c|}{0.1}          & 1            & \multicolumn{1}{c|}{}                       & \multicolumn{1}{c|}{}                        & \multicolumn{1}{c|}{}                       &                      \\ \cline{2-6}
                            & \multicolumn{1}{c|}{CD}              & \multicolumn{1}{c|}{-4.5}         & \multicolumn{1}{c|}{1.4}          & \multicolumn{1}{c|}{1}            & 0.5          & \multicolumn{1}{c|}{}                       & \multicolumn{1}{c|}{}                        & \multicolumn{1}{c|}{}                       &                      \\ \hline
\multirow{3}{*}{Scenario F} & \multicolumn{1}{c|}{CA}              & \multicolumn{1}{c|}{-6.36}        & \multicolumn{1}{c|}{1.5035}       & \multicolumn{1}{c|}{1.1}          & 0.5          & \multicolumn{1}{c|}{\multirow{3}{*}{-6.36}} & \multicolumn{1}{c|}{\multirow{3}{*}{1.5035}} & \multicolumn{1}{c|}{\multirow{3}{*}{1.1}}   & \multirow{3}{*}{0.5} \\ \cline{2-6}
                            & \multicolumn{1}{c|}{PA}              & \multicolumn{1}{c|}{-6.36}        & \multicolumn{1}{c|}{1.65}         & \multicolumn{1}{c|}{1.1}          & 0.5          & \multicolumn{1}{c|}{}                       & \multicolumn{1}{c|}{}                        & \multicolumn{1}{c|}{}                       &                      \\ \cline{2-6}
                            & \multicolumn{1}{c|}{CD}              & \multicolumn{1}{c|}{-1.5}         & \multicolumn{1}{c|}{-1}           & \multicolumn{1}{c|}{-1}           & 0.25         & \multicolumn{1}{c|}{}                       & \multicolumn{1}{c|}{}                        & \multicolumn{1}{c|}{}                       &                      \\ \hline
\multirow{3}{*}{Scenario G} & \multicolumn{1}{c|}{CA}              & \multicolumn{1}{c|}{-6}           & \multicolumn{1}{c|}{1.1}          & \multicolumn{1}{c|}{1.323}        & 0            & \multicolumn{1}{c|}{\multirow{3}{*}{-6}}    & \multicolumn{1}{c|}{\multirow{3}{*}{1.1}}    & \multicolumn{1}{c|}{\multirow{3}{*}{1.323}} & \multirow{3}{*}{0}   \\ \cline{2-6}
                            & \multicolumn{1}{c|}{PA}              & \multicolumn{1}{c|}{-7}           & \multicolumn{1}{c|}{1.1}          & \multicolumn{1}{c|}{1.4}          & 0            & \multicolumn{1}{c|}{}                       & \multicolumn{1}{c|}{}                        & \multicolumn{1}{c|}{}                       &                      \\ \cline{2-6}
                            & \multicolumn{1}{c|}{CD}              & \multicolumn{1}{c|}{-6}           & \multicolumn{1}{c|}{1.622}        & \multicolumn{1}{c|}{1.2}          & 0            & \multicolumn{1}{c|}{}                       & \multicolumn{1}{c|}{}                        & \multicolumn{1}{c|}{}                       &                      \\ \hline
\multirow{3}{*}{Scenario H} & \multicolumn{1}{c|}{CA}              & \multicolumn{1}{c|}{-5.35}        & \multicolumn{1}{c|}{1.025}        & \multicolumn{1}{c|}{0.7}          & 3            & \multicolumn{1}{c|}{\multirow{3}{*}{-5.35}} & \multicolumn{1}{c|}{\multirow{3}{*}{1.025}}  & \multicolumn{1}{c|}{\multirow{3}{*}{0.7}}   & \multirow{3}{*}{3}   \\ \cline{2-6}
                            & \multicolumn{1}{c|}{PA}              & \multicolumn{1}{c|}{-5}           & \multicolumn{1}{c|}{1.1}          & \multicolumn{1}{c|}{-1}           & 1            & \multicolumn{1}{c|}{}                       & \multicolumn{1}{c|}{}                        & \multicolumn{1}{c|}{}                       &                      \\ \cline{2-6}
                            & \multicolumn{1}{c|}{CD}              & \multicolumn{1}{c|}{-3.54}        & \multicolumn{1}{c|}{0.5}          & \multicolumn{1}{c|}{1}            & 1            & \multicolumn{1}{c|}{}                       & \multicolumn{1}{c|}{}                        & \multicolumn{1}{c|}{}                       &                      \\ \hline
\end{tabular}
\label{table_efficacy_scenarios_H0}
\end{table}

\begin{table}[H]
\caption{Average sample size under the early stopping for futility rule under $H_1$ (Scenarios A-D) and $H_0$ (Scenarios E-H).}
\centering
\vspace{0.25cm}
\begin{tabular}{|c|ccc|ccc|ccc|ccc|}
\hline
\multirow{2}{*}{} & \multicolumn{3}{c|}{Scenario A}                        & \multicolumn{3}{c|}{Scenario B}                        & \multicolumn{3}{c|}{Scenario C}                        & \multicolumn{3}{c|}{Scenario D}                        \\ \cline{2-13} 
                  & \multicolumn{1}{c|}{CA} & \multicolumn{1}{c|}{PA} & CD & \multicolumn{1}{c|}{CA} & \multicolumn{1}{c|}{PA} & CD & \multicolumn{1}{c|}{CA} & \multicolumn{1}{c|}{PA} & CD & \multicolumn{1}{c|}{CA} & \multicolumn{1}{c|}{PA} & CD \\ \hline
$\omega = 0$      & \multicolumn{1}{c|}{53} & \multicolumn{1}{c|}{53} & 53 & \multicolumn{1}{c|}{55} & \multicolumn{1}{c|}{55} & 55 & \multicolumn{1}{c|}{53} & \multicolumn{1}{c|}{53} & 53 & \multicolumn{1}{c|}{55} & \multicolumn{1}{c|}{55} & 55 \\ \hline
$\omega = 0.25$   & \multicolumn{1}{c|}{53} & \multicolumn{1}{c|}{53} & 54 & \multicolumn{1}{c|}{55} & \multicolumn{1}{c|}{55} & 55 & \multicolumn{1}{c|}{53} & \multicolumn{1}{c|}{53} & 53 & \multicolumn{1}{c|}{55} & \multicolumn{1}{c|}{55} & 55 \\ \hline
$\omega = 0.5$    & \multicolumn{1}{c|}{54} & \multicolumn{1}{c|}{54} & 54 & \multicolumn{1}{c|}{55} & \multicolumn{1}{c|}{55} & 55 & \multicolumn{1}{c|}{54} & \multicolumn{1}{c|}{54} & 54 & \multicolumn{1}{c|}{55} & \multicolumn{1}{c|}{55} & 55 \\ \hline
$\omega = 0.75$   & \multicolumn{1}{c|}{54} & \multicolumn{1}{c|}{54} & 54 & \multicolumn{1}{c|}{55} & \multicolumn{1}{c|}{55} & 55 & \multicolumn{1}{c|}{54} & \multicolumn{1}{c|}{54} & 54 & \multicolumn{1}{c|}{55} & \multicolumn{1}{c|}{55} & 55 \\ \hline
$\omega = 1$      & \multicolumn{1}{c|}{54} & \multicolumn{1}{c|}{54} & 54 & \multicolumn{1}{c|}{55} & \multicolumn{1}{c|}{55} & 55 & \multicolumn{1}{c|}{54} & \multicolumn{1}{c|}{54} & 54 & \multicolumn{1}{c|}{55} & \multicolumn{1}{c|}{55} & 55 \\ \hline
\multirow{2}{*}{} & \multicolumn{3}{c|}{Scenario E}                        & \multicolumn{3}{c|}{Scenario F}                        & \multicolumn{3}{c|}{Scenario G}                        & \multicolumn{3}{c|}{Scenario H}                        \\ \cline{2-13} 
                  & \multicolumn{1}{c|}{CA} & \multicolumn{1}{c|}{PA} & CD & \multicolumn{1}{c|}{CA} & \multicolumn{1}{c|}{PA} & CD & \multicolumn{1}{c|}{CA} & \multicolumn{1}{c|}{PA} & CD & \multicolumn{1}{c|}{CA} & \multicolumn{1}{c|}{PA} & CD \\ \hline
$\omega = 0$      & \multicolumn{1}{c|}{45} & \multicolumn{1}{c|}{45} & 45 & \multicolumn{1}{c|}{46} & \multicolumn{1}{c|}{46} & 46 & \multicolumn{1}{c|}{46} & \multicolumn{1}{c|}{46} & 46 & \multicolumn{1}{c|}{47} & \multicolumn{1}{c|}{47} & 47 \\ \hline
$\omega = 0.25$   & \multicolumn{1}{c|}{46} & \multicolumn{1}{c|}{46} & 46 & \multicolumn{1}{c|}{47} & \multicolumn{1}{c|}{47} & 47 & \multicolumn{1}{c|}{47} & \multicolumn{1}{c|}{47} & 47 & \multicolumn{1}{c|}{48} & \multicolumn{1}{c|}{48} & 49 \\ \hline
$\omega = 0.5$    & \multicolumn{1}{c|}{46} & \multicolumn{1}{c|}{46} & 47 & \multicolumn{1}{c|}{47} & \multicolumn{1}{c|}{47} & 47 & \multicolumn{1}{c|}{47} & \multicolumn{1}{c|}{47} & 48 & \multicolumn{1}{c|}{48} & \multicolumn{1}{c|}{48} & 48 \\ \hline
$\omega = 0.75$   & \multicolumn{1}{c|}{46} & \multicolumn{1}{c|}{46} & 47 & \multicolumn{1}{c|}{48} & \multicolumn{1}{c|}{48} & 47 & \multicolumn{1}{c|}{48} & \multicolumn{1}{c|}{47} & 48 & \multicolumn{1}{c|}{49} & \multicolumn{1}{c|}{49} & 49 \\ \hline
$\omega = 1$      & \multicolumn{1}{c|}{47} & \multicolumn{1}{c|}{47} & 47 & \multicolumn{1}{c|}{48} & \multicolumn{1}{c|}{48} & 47 & \multicolumn{1}{c|}{48} & \multicolumn{1}{c|}{48} & 48 & \multicolumn{1}{c|}{49} & \multicolumn{1}{c|}{49} & 49 \\ \hline
\end{tabular}
\end{table}

\begin{figure}[H]
\caption{Probability of rejecting $H_0$ in scenarios under $H_1$ (i.e., power) and $H_0$ (i.e., type-I error). Scenario A-D and E-H are scenario under $H_1$ and $H_0$, respectively.}
\centering
\vspace{0.25cm}
\includegraphics[scale=0.7]{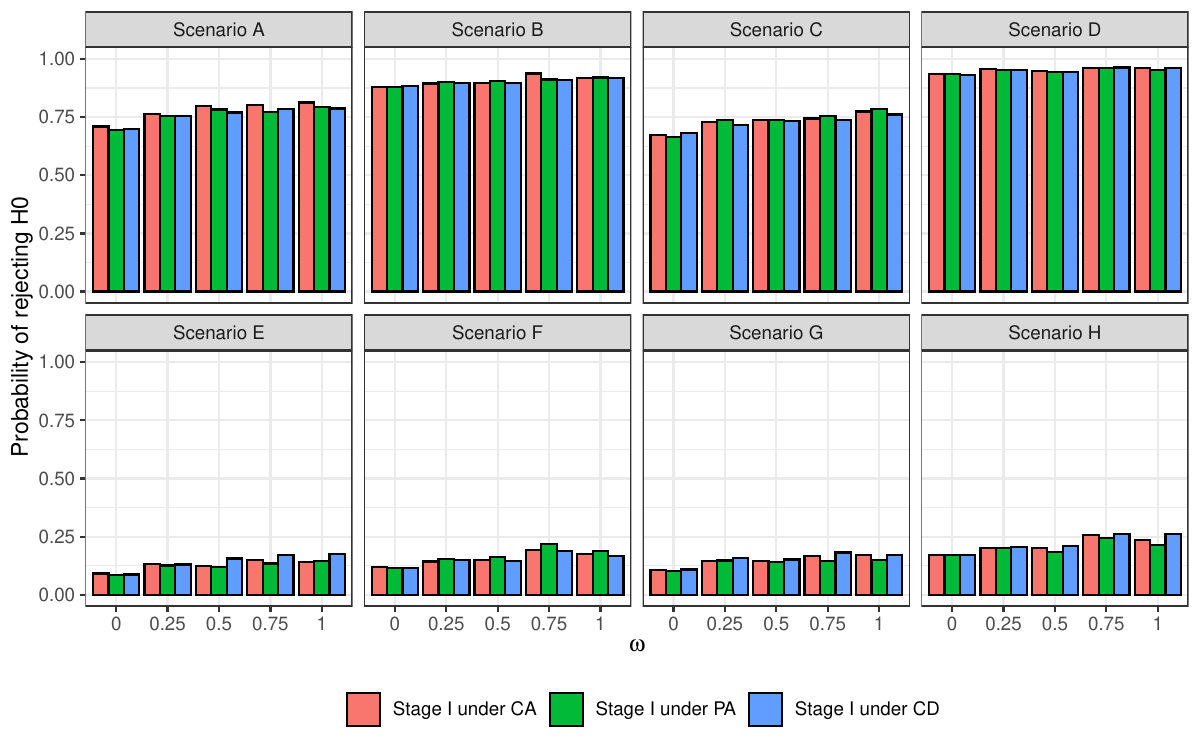}
\label{Figure_S1}
\end{figure}

\begin{figure}[H]
\caption{Difference in the proportion of patients allocated to dose combination with true probability of efficacy above $p_0$ (i.e., patients correctly allocated) in stage II with $\omega > 0$ with respect to $\omega = 0$.}
\centering
\vspace{0.25cm}
\includegraphics[scale=0.7]{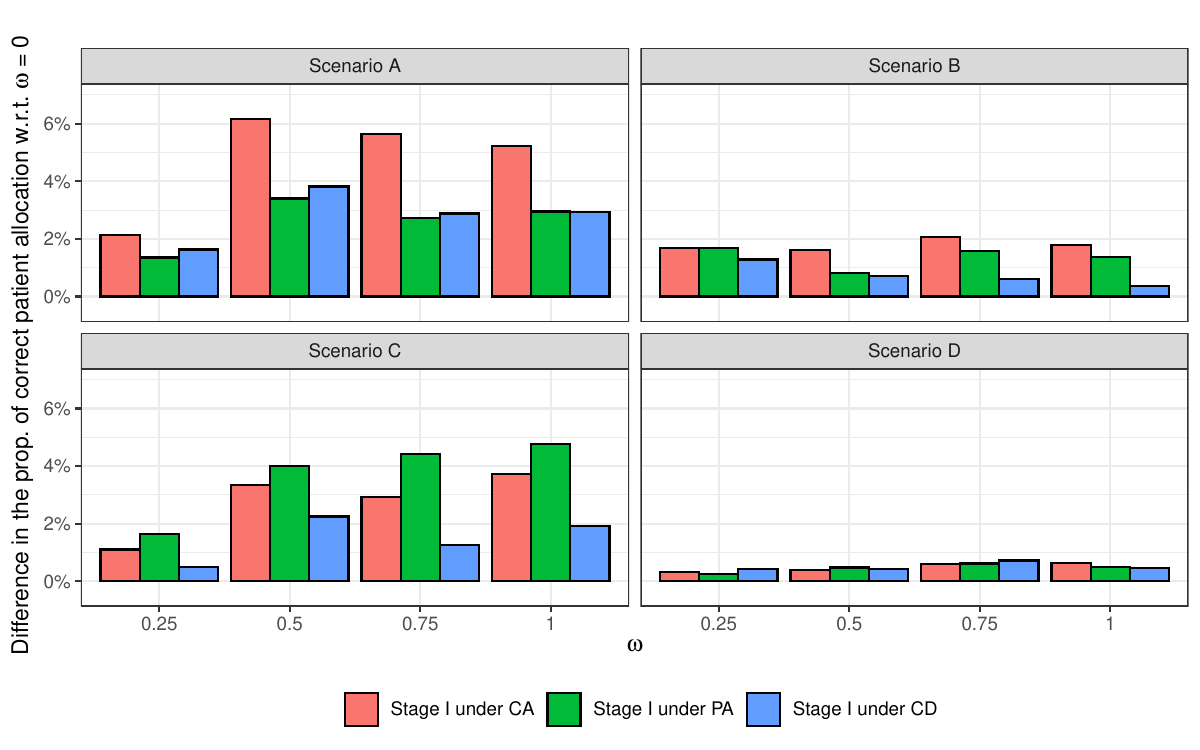}
\label{Figure_S2}
\end{figure}

\begin{figure}[H]
\caption{Probability of early stopping for safety.}
\centering
\vspace{0.25cm}
\includegraphics[scale=0.7]{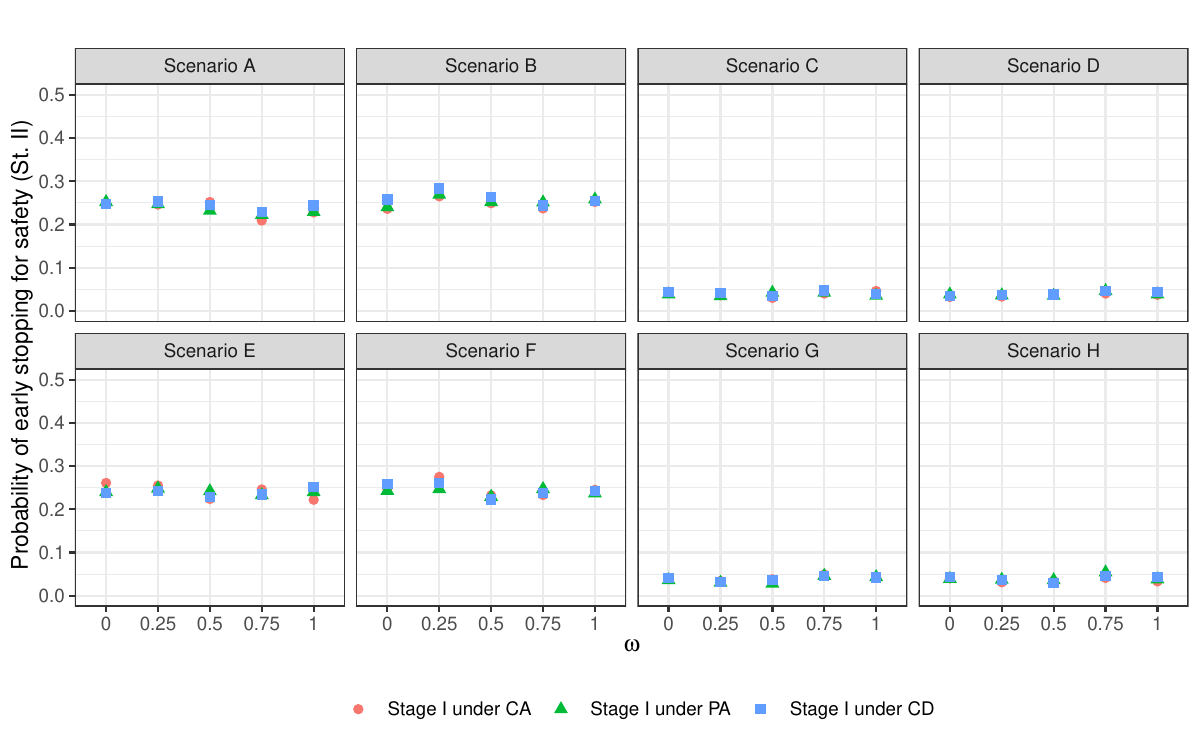}
\label{Figure_S5}
\end{figure}

\end{document}